\documentclass[english,aps,twocolumn,showpacs,amsmath,amssymb,prx,footinbib,superscriptaddress]{revtex4-2}

\usepackage[T1]{fontenc}
\usepackage[utf8]{inputenc}
\makeatletter
\usepackage{lineno}
\usepackage{graphicx}
\usepackage{xcolor}
\usepackage{amsmath}
\usepackage{psfrag}
\usepackage[abs]{overpic}
\usepackage{enumitem}
\usepackage{natbib}
\usepackage{float}
\usepackage{verbatim}
\usepackage{soul}
\usepackage[normalem]{ulem}
\usepackage{babel}
\usepackage{fancyhdr}
\usepackage[%
colorlinks=true,
urlcolor=blue,
linkcolor=blue,
citecolor=blue
]{hyperref}

\newcommand{\vctr}[1]{\boldsymbol{\mathrm{#1}}}
\newcommand{\mtrx}[1]{\underline{\mathrm{#1}}}

\makeatother

\begin{document}

\title{Quasiperiodic circuit quantum electrodynamics}

\author{T. Herrig}
\affiliation{Peter Gr\"unberg Institute, Theoretical Nanoelectronics, Forschungszentrum J\"ulich, D-52425 J\"ulich, Germany}

\author{J. H. Pixley}
\affiliation{Department of Physics and Astronomy, Center for Materials Theory, Rutgers University, Piscataway, New Jersey 08854, USA}
\affiliation{Center for Computational Quantum Physics, Flatiron Institute, 162 5th Avenue, New York, NY 10010}

\author{E. J. K\"onig}
\email{e.koenig@fkf.mpg.de}
\affiliation{Max-Planck Institute for Solid State Research, 70569 Stuttgart, Germany}

\author{R.-P. Riwar}
\affiliation{Peter Gr\"unberg Institute, Theoretical Nanoelectronics, Forschungszentrum J\"ulich, D-52425 J\"ulich, Germany}

\begin{abstract}
    Superconducting circuits are an extremely versatile platform to realize quantum information hardware and to emulate topological materials. We here show how a simple arrangement of capacitors and conventional superconductor-insulator-superconductor junctions can realize an even broader class of systems, in the form of a nonlinear capacitive element which is quasiperiodic with respect to the quantized Cooper-pair charge. Our setup allows to create protected Dirac points defined in the transport degrees of freedom, whose presence leads to a suppression of the classical finite-frequency current noise. Furthermore, the quasiperiodicity can emulate Anderson localization in charge space, measurable via vanishing charge quantum fluctuations. The realization by means of the macroscopic transport degrees of freedom allows for a straightforward generalization to arbitrary dimensions and implements truly non-interacting versions of the considered models. As an outlook, we discuss potential ideas to simulate a transport version of the magic-angle effect known from twisted bilayer graphene.
\end{abstract}

\maketitle

\section*{Introduction}

Superconducting circuits are a prime candidate for the realization of large scale quantum computers~\cite{Arute_2019_short,IBM_roadmap}. Within this thrust, it has been noticed in the last years that these circuits harbor an enormous potential for the realization of topological phases of matter, surprisingly without need for topological or strongly correlated materials~\cite{Leone2008,Leone_2013,Yokoyama2015TopolABSmultitJJ,Riwar2016,Strambini_2016,Eriksson2017,Meyer:2017aa,Xie:2017aa,Xie:2018aa,Deb:2018aa,Repin:2019aa,Repin_2020,Fatemi_2020,Peyruchat_2020,Klees:2020aa,Klees_2021,Weisbrich_2021,Weisbrich_2021_monopoles,Chirolli_2021,Herrig2022,Melo2022}. In a nutshell, the idea is that instead of considering the regular band structure of a material (obtained from position and momentum degrees of freedom of the electron), the transport degrees of freedom of the circuit (charge and phase across a given circuit branch) may encode a given topological invariant.
	
However, when it comes to the emulation of condensed matter systems by means of superconducting circuits there remains until now a huge patch of uncharted territory: quasiperiodicity. In solid state systems, quasiperiodicity may appear on the mean field level in systems with incommensurate charge density or spin density waves. Moreover, and possibly more importantly for applications, incommensurate lattices ubiquitously appear in heterostructures of van-der-Waals materials~\cite{GeimGrigorieva2913}. Within this context, twisted heterostructures~\cite{CaoJarillo} at incommensurate twist-angles~\cite{FuPixley2020,GonzalezCirac2019,SalamonRakshit2020,ChouPixley2020,MaoSenthil2021,MengZhang2021,LeePixley2022} play a particularly interesting role. Finally, electrons moving in a quasicrystalline environment are subject to incommensuration effects, which can lead to emergent critical behavior~\cite{GoldYbAl}. 
Motivated by these examples, we here intend to show how quasiperiodicity can be engineered in the transport degrees of freedom of a circuit by very straightforward means, and how it can be exploited to unlock previously inaccessible circuit behavior. The key to this endeavor turns out to be a peculiar form of a nonlinear capacitor.

By far the most common capacitor is the regular, linear capacitor, whose energy depends quadratically on the charge $N$
\begin{equation}
		E=\frac{2e^2}{C} N^2\,,
\end{equation}
where $C$ is the capacitance, and the charge number $N$ is counted in units of Cooper-pairs. The quadratic form is simply a consequence of standard electrodynamics, where the energy stored in a capacitor is computed by means of the square of the electric field. Now, it is interesting to think about alternative capacitances that are, in one way or another, nonlinear. Ferro-electric materials~\cite{Landauer_1976,Catalan_2015,Ng_2017,Hoffmann_2018,Lukyanchuk_2019,Hoffmann_2020}, for instance, provide a mechanism leading to a highly nonlinear capacitive behavior, with an energy which can locally (close to zero charge on the capacitor) indeed be approximated yet again as quadratic, but suprisingly, with an effectively \textit{negative} capacitance, $C<0$. While the total capacitance of any charged island must of course be positive to guarantee a lower bound of the Hamiltonian eigenenergies, partial negative capacitances are a real phenomenon, and have, e.g., been proposed as a lever arm to amplify the voltage sensitivity of field effect transistors~\cite{Salahuddin_2008}. Another way to engineer the electrostatic properties of an island is a capacitive coupling to a nearby electronic quantum system, a so-called polarizer, whose eigenenergy depends yet again on the island charge in a nontrivial way. This idea was pioneered by Little~\cite{Little_1964} to induce superconducting pairing in the absence of electron-phonon interactions. It has recently been studied both experimentally and theoretically in quantum dot systems to realize attractive interactions~\cite{Hamo_2016,Placke_2018}---yet again a form of negative capacitance. Finally, effective negative capacitances have recently been shown to occur also as a dynamic effect in the course of a time-dependent flux-drive, due to surface charges induced by the electro-motive force~\cite{Riwar_2022}.

Here, we propose a circuit which realizes a \textit{quasiperiodic} nonlinear capacitor (QPNC) by coupling to a nearby transmon. The resulting energy term is of the form
\begin{equation}\label{eq:intro-quasiperiodic-energy}
	E = - E_\text{S} \cos\left(2\pi\lambda N\right),
\end{equation}
with the real parameter $\lambda$ determining the periodicity of the capacitance. At the surface, this seems to generalize the nonlinear capacitance of phase slip junctions~\cite{Buchler2004,Mooij_2006,Astafiev_2012,deGraaf_2018} to a term whose periodicity we can choose by design. Note however, that contrary to conventional phase slip junctions where charge quantization enters in a different form~\cite{koliofoti2022}, in our case, the QPNC is compatible with $N\in \mathbb{Z}$. Interestingly, $\lambda$ can assume any real value since it is determined by a ratio of capacities; therefore it will naturally assume an irrational value. In conjunction with $N\in \mathbb{Z}$, this leads to the capacitance being in general quasiperiodic in charge space.

To illustrate the versatility of this circuit element, we explore various consequent effects in different combinations with other elements. In fact, the possibility to realize a nonlinear capacitance greatly expands the toolbox of building blocks to simulate condensed matter systems.
To start, we propose a setup which emulates a Dirac material. The topology is defined in a mixed circuit-parameter space of a superconducting phase and an offset charge, which was already studied earlier to provide a highly stable circuit realization of a Chern insulator~\cite{Herrig2022}. The presence of the Dirac points has a measurable influence on the current noise spectrum. In particular, we show that the topology-induced vanishing of the Berry curvature will suppress the influence of classical finite-frequency charge noise.

In a different setup, we fully utilize the quasiperiodicity of the nonlinear capacitive energy term by realizing a simulation of the Aubry-Andr\'e (AA) model~\cite{AubryAndre,Harper_1955}, a paradigmatic lattice model for Anderson localization of quantum particles moving in a quasiperiodic background potential. Our truly non-interacting setup yields a platform which allows to directly probe Anderson localization in charge space.
The AA model is known to display a quantum phase transition from extended to Anderson-localized states and has attracted enormous attention over the years and in various fields of research. It represents a fruit fly model of localization which is solvable using rigorous mathematical~\cite{TenMartini} and mathematical physics methods~\cite{WiegmannZabrodin} and has been implemented in a variety of experimental setups, including polaritonic waveguides~\cite{Goblot:2020aa} as well as cold atomic gases~\cite{RoatiInguscio2008}. It is moreover related to the Hofstadter-butterfly problem of 2D Bloch electrons in a magnetic field~\cite{Hofstadter_1976}. Clearly, different experimental emulators of the AA model suffer distinctly from a variety of imperfections (e.g.\ particle loss, finite size effects, residual interactions, and associated reduced quantum coherence at finite temperature). The proposed superconducting circuit offers a platform which in particular mitigates the problem of residual interactions: It is based on the coherent macroscopic quantum mechanics of superconducting circuits in which microscopic interactions are already fully incorporated. 

A major advantage of the proposed set up is its versatility. First, we show how this system can emulate the low energy excitations of a two-dimensional Dirac semimetal. We then demonstrate how this setup can also be used to realize the AA model as well as its generalization to higher dimensions~\cite{Sokoloff1980,DevakulHuse2017,Luo-2022}. The nature of quasiperiodic localization in higher dimensions and whether or not it is distinct from its random counterpart in any way remains a timely question and our proposed set up represents an experimental test of this. Last, we discuss how to combine these perspectives to realize an emulator of the magic-angle phenomena in twisted bilayer graphene~\cite{FuPixley2020,GonzalezCirac2019, SalamonRakshit2020, LeePixley2022}.

\section*{results}

\subsection*{Quasiperiodic capacitance} \label{sec:quasiperiodic-capacitance}

\begin{figure*}
	\centering
	\includegraphics[width=.75\linewidth]{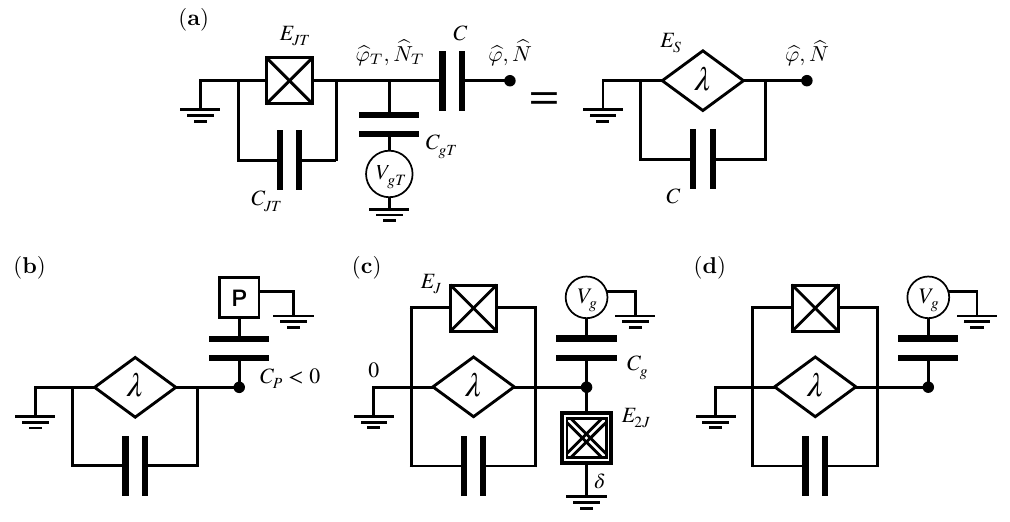}
	
	\caption{\textbf{Circuit realizing a quasiperiodic nonlinear capacitor (QPNC) and applications.} (a) The element is implemented by a fast transmon capacitively coupled (with capacity $C$) to a superconducting island (with phase $\widehat{\varphi}$ and charge $\widehat{N}$) at the right contact. That is, we have an auxiliary island for the transmon (with phase $\widehat{\varphi}_\text{T}$ and charge $\widehat{N}_\text{T}$) which is connected to ground via a Josephson junction (of energy $E_\text{JT}$ and associated with a capacity $C_\text{JT}$). For completeness, we also include a gate voltage $V_\text{gT}$ (via a capacitance $C_\text{gT}$). The right island is assumed to be connected to further elements as part of a larger circuit, like in (c, d). The QPNC is associated with an energy scale $E_\text{S}$ and the quasiperiodicity parameter $\lambda$, and it always comes with a parasitic linear capacitance with the same capacity $C$ as the coupling.
	(b) One way to counteract the parasitic linear capacitance is by making use of a \emph{partial} negative capacitance, here realized by a polarizer (P) giving an energy contribution effectively described by a negative capacity $C_\text{P} < 0$.
	(c) The Dirac circuit realizes topologically protected degeneracies in 2D. The QPNC is placed in parallel with a Josephson junction (of energy $E_\text{J}$) and in series with a cotunneling junction (of energy $E_\text{2J}$) connecting to ground with a phase shift of $\delta$. A cotunneling junction only allows for pairs of Cooper-pairs to tunnel. A gate voltage $V_\text{g}$ (via a capacitance $C_\text{g}$) is also added to the island.
	(d) The Aubry-Andr\'e circuit is the same as the Dirac circuit but without the cotunneling junction. It simulates a trapped version of the Aubry-Andr\'e model on a charge lattice.
	}
	\label{fig:master-figure}
\end{figure*}

Consider an island with Cooper-pair number operator $\widehat N$ and superconducting phase $\varphi$. Let us add a capacitively coupled auxiliary transmon, with Cooper-pair number operator $\widehat{N}_\text{T}$ and a Josephson junction of energy $E_\text{JT}$ coupling to the ground; see left circuit of Fig.~\ref{fig:master-figure}(a). The offset charge $N_\text{gT} = C_\text{gT} V_\text{gT} / 2e$ is controlled via a gate voltage $V_\text{gT}$. 
We remind the reader that this is not the complete circuit but just a subsystem of a larger circuit, such that eventually, more elements will be connected to the contacts (giving rise to a dynamics of the island degrees of freedom). This subsystem, however, effectively realizes a quasiperiodic nonlinear capacitor (QPNC), as we will see below.

The Hamiltonian of the subsystem reads
\begin{multline}\label{eq:detailed-QPNC}
	\widehat{H} = 2E_\text{CT} \left(\widehat{N}_\text{T} + N_\text{gT} + \lambda \widehat{N}\right)^{2}\\
	+ 2E_\text{C} \widehat{N}^2 - E_\text{JT} \cos(\widehat{\varphi}_\text{T})\,,
\end{multline}
where the superconducting phase operator $\widehat{\varphi}_\text{T}$ is canonically conjugate to the number operator $\widehat{N}_\text{T}$, such that $\bigl[\widehat{\varphi}_\text{T}, \widehat{N}_\text{T}\bigr] = \mathrm{i}$. Find a detailed derivation of this Hamiltonian in the Supplementary Note 1.
The first two terms of the Hamiltonian are the charging energies of the island and the transmon, $E_\text{C} = e^2 / C_\text{tot}$ and $E_\text{CT} = e^2 C_\text{tot} / (C_\text{tot} C_{\text{T,tot}} - C^2)$, with the total capacities of the island and transmon, here, $C_\text{tot} = C$ and $C_{\text{T,tot}} = C + C_\text{gT} + C_\text{JT}$. The parameter
\begin{equation}\label{eq:lambda}
	\lambda = C / C_\text{tot}
\end{equation} 
will play the crucial role of determining the (quasi)periodicity of the nonlinear capacitance. Note importantly that as of now, $\lambda=1$, since the only capacitance the island has is the one which is coupling to the transmon.  In a complete model, $\lambda$ will not be trivially one, as Eq.~\eqref{eq:lambda} implies here, but instead, the total capacitance of the island $C_\text{tot}$ will differ from $C$ due to additional capacitive elements; see Supplementary Note 1. That is, in general, $\lambda$ expresses the partition of the island charge that couples to the transmon.
This is the reason why we deliberately chose to write the parameters ($E_\text{C}$, $E_\text{CT}$, and $\lambda$) in the above non-simplified form, that is, to generalize them to arbitrary applications in larger circuits with $C_\text{tot} \neq C$.

Note that we could also invert the QPNC, that is, starting from Fig.~\ref{fig:master-figure}(a), we switch the island at the right contact with the ground at the left contact, and thus the charges $\widehat{N}$ and $\widehat{N}_\text{T}$ are inductively connected while only capacitively coupled to ground. While this changes a few details in the treatment, the actual effects stay the same when the auxiliary transmon island only couples to a single other island (and to ground). However, one additional effect is that $\lambda$ is mapped to negative values, which becomes relevant when extending the model by capacitively coupling additional islands to the same QPNC as we show below.

In the following we assume that the dynamics of the auxiliary transmon are much faster than any dynamics of $\widehat{N}$. Thus we first solve the transmon Hamiltonian while leaving $N$ constant. This allows us to simplify the Hamiltonian given in Eq.~\eqref{eq:detailed-QPNC} to the low-energy Hamiltonian, where we only take into account the lowest energy level of the transmon~\cite{Cottet_2002}
\begin{equation}\label{eq:eff-qp-Ham}
	\widehat{H} \approx -E_\text{S} \cos\left(2\pi \lambda \widehat{N} + 2\pi N_\text{gT}\right) + 2E_\text{C} \widehat{N}^2\ .
\end{equation}
We identify a nonlinear capacitance in addition to an unavoidable linear capacitance, such that we can express the circuit as two parallel capacitive elements, as shown in Fig.~\ref{fig:master-figure}(a). Strictly speaking, the above low-energy approximation is well-justified in the transmon regime $E_\text{CT} < E_\text{JT}$, where the amplitude
\begin{equation}\label{eq:E_S}
	E_\text{S} = 16 \left(\frac{E_\text{CT} E_\text{JT}^{3}}{4 \pi^2}\right)^\frac{1}{4} \mathrm{e}^{-4 \sqrt{\frac{E_\text{JT}}{E_\text{CT}}}}\ ,
\end{equation}
is small and the energy gap to the first excited state of the transmon $\sim 2\sqrt{E_\text{CT} E_\text{JT}}$ is large. However, for our purposes it is advantageous to have a sufficiently large $E_\text{S}$ such that its effect is not merely perturbative. We therefore note that we do not necessarily need to go very deep into the transmon regime, where $E_\text{CT}\ll E_\text{JT}$: the cosine expression is still a good approximation when we consider the intermediate regime of $E_\text{CT} \sim E_\text{JT}$, where $E_\text{S}$ is not exponentially suppressed; see Supplementary Note 2. Instead we find $E_\text{S}$ and $E_\text{CT}$ to have a similar order of magnitude. To be precise, the value of $E_\text{S}$ maxes out at $\sim E_\text{CT}/4$, which can be easily seen by looking at the Hamiltonian in Eq.~\eqref{eq:detailed-QPNC} in the limit $E_\text{JT} \rightarrow 0$. A very minor drawback of $E_\text{CT}\sim E_\text{JT}$ is that the energy gap to the first excited level of the transmon is no longer large, such that it can be occupied. Since the energy dependence of the excited level resembles a cosine with an amplitude of reversed sign, these excitations effectively reduce again $E_\text{S}$, such that there is an optimal trade-off for moderate parameter regimes; for more details see Supplementary Note 3. At any rate, we identify the regime in which our low-energy description is valid as the one where $E_\text{C}$ does not exceed the gap. For $E_\text{CT}\sim E_\text{JT}$ that provides the bound $E_\text{C} \lesssim E_\text{CT}$.

A remaining obstacle that we have to discuss is the possible parameter range for the ratio between $E_\text{S}$ and $E_\text{C}$. For the nonlinear capacitor effect to become dominant, we would ideally need $E_\text{S} \gg E_\text{C}$. Importantly however, these two energies are not independent, as both depend rather non-trivially on the capacitances of the circuit. As a matter of fact, if all capacitances involved are regular, positive capacitances, our analysis shows that $E_\text{S}\gg E_\text{C}$ can only be realized for $\lambda$ close to one or close to zero, which might seem to severely limit the applicability of our proposal. But there are several successful work-arounds. One possibility we explore in this work is to selectively suppress $E_\text{C}$ (but not $E_\text{S}$) by adding nontrivial Josephson junction elements. Another strategy is to allow for partial negative capacitances. Let us first demonstrate that negative capacitances actually lift any remaining restrictions, such that we can tune $E_\text{S}/E_\text{C}$ and $\lambda$ at will. Then we briefly sketch some feasible ideas for implementing negative capacitances based on the polarizer principle~\cite{Little_1964,Hamo_2016,Placke_2018}.

\begin{figure}
	\centering
	\includegraphics[width=.8\linewidth]{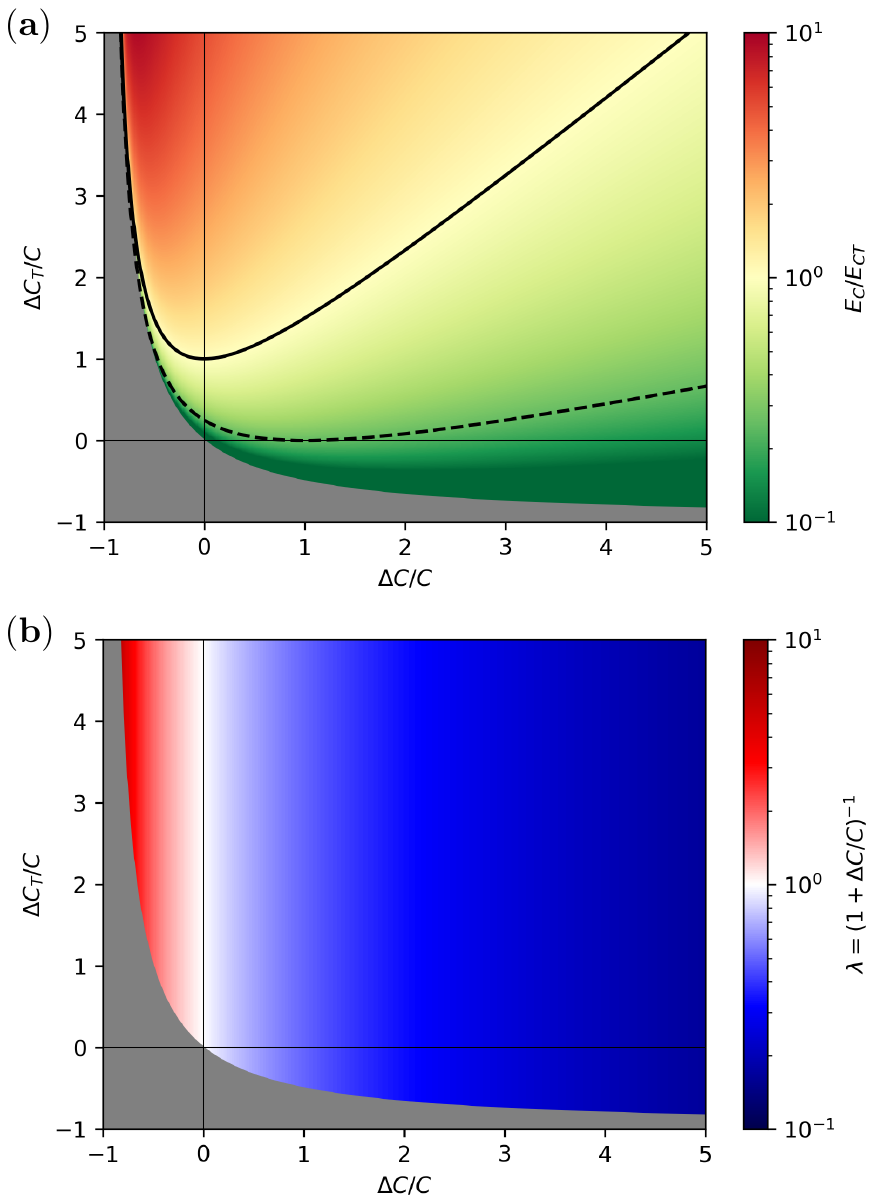}
	
	\caption{\textbf{Evaluation of allowed parameter regimes to achieve a dominant quasiperiodic capacitance.} Depicted are the achievable parameter values for (a) the ratio $E_\text{C} / E_\text{CT}$ and (b) $\lambda$ as a function of the capacitances of the superconducting islands. The grey areas are illegal instability-regions marked by a negative charging energy. Note that we could also achieve negative values for $\lambda$ when choosing $C$ to be negative. (a) The values below $E_\text{C} < 0.1 E_\text{CT}$ are cut off and depicted in the same shade of green. The solid black line marks the border for which $E_\text{C} = E_\text{CT}$, thus, where the low-energy approximation breaks down. The dashed line is at a value of $E_\text{C} / E_\text{CT} = 1 / 4$ under which $E_\text{S}$ can exceed $E_\text{C}$.
	}
	\label{fig:negative-capacitances}
\end{figure}

In Fig.~\ref{fig:negative-capacitances} we choose to depict the ratio $E_\text{C} / E_\text{CT}$ as well as $\lambda$ as a function of $\Delta C \equiv C_{\text{tot}} - C$ and $\Delta C_\text{T} \equiv C_{\text{T,tot}} - C$ relative to $C$. The ratio $E_\text{C}/E_\text{CT}$ is readily translated to $\sim E_\text{C} / 4 E_\text{S}$, see the discussion above on the maximal value for $E_\text{S}$ and, for more details, Supplementary Note 2. Note that we allow for partial negative capacitances (i.e., $\Delta C$ and $\Delta C_\text{T}$ may be negative), but only insofar as $E_\text{C},E_\text{CT} > 0$ (that is, $C_\text{tot},C_{\text{T,tot}} > 0$), as otherwise there is no lower bound in the Hamiltonian (as stated already in the introduction). Illegal values for the capacitances, which break the positivity of the charging energies, are greyed out. Note that we can also consider $C$ to be negative (with the same restrictions on positivity of course), which yields a mirror image of Fig.~\ref{fig:negative-capacitances} but with negative values for $\lambda$, without the need for inverting the QPNC. However, the below presented polarizer concept for negative capacitances cannot realize negative cross-couplings between two islands, which would require methods like using ferro-electric materials.
The values of $\Delta C$ and $\lambda$ are directly related [see Fig.~\ref{fig:negative-capacitances}(b)], which is obvious, given the relationship in Eq.~\eqref{eq:lambda}. The impact of $\Delta C_\text{T}$ on the other hand requires a more detailed discussion. As a general rule, we find that $\Delta C_\text{T}$ should be as small as possible; see Fig.~\ref{fig:negative-capacitances}(a). The solid line marks $E_\text{C} = E_\text{CT}$ that is the border above which the low-energy approximation breaks down. More importantly, $E_\text{C}$ can drop below $E_\text{S}\sim E_\text{CT} / 4$ only for values underneath the dashed line. When using strictly positive capacitances, $E_\text{C} / E_\text{CT}$ has a minimal reachable value depending on $\lambda$ which is the highest for $\lambda = 1/2$ at $E_\text{C} / E_\text{CT} = 1/4$. This restriction is only lifted close to the trivial value $\lambda = 1$, or for general $\lambda$ when allowing for partial negative capacitances.

As already stated in the introduction, ferro-electric materials~\cite{Landauer_1976,Catalan_2015,Ng_2017,Hoffmann_2018,Lukyanchuk_2019,Hoffmann_2020} can be used to create partial negative capacitances. However, as of today, we are unaware of attempts to include ferro-electric materials into superconducting devices at low temperatures. We believe however, that there is a much more feasible approach, readily integrable into superconducting device architecture, relying on a capacitive coupling to a polarizer. Ultimately, the effect of the polarizer is included in exactly the same way as for the auxiliary transmon [for a schematic, see Fig.~\ref{fig:master-figure}(b)]. We again fix the island charge $N$, and compute the ground state energy of the polarizer as a function of $N$ (assuming again the existence of an energy gap, see below). This energy is added as a contribution to the total charging energy of the resulting circuit. To illustrate the basic principle studied by Ref.~\cite{Placke_2018}, consider a capacitive coupling between the island and a generic two-level system, with a Hamiltonian of the form
\begin{equation}
    \widehat{H}_\text{P} = u N \widehat{\sigma}_z + p \widehat{\sigma}_x\,.
\end{equation}
The resulting eigenvalues are $\pm \sqrt{u^2 N^2 +p^2}$. As indicated above, assuming a large gap (here given by $p$), we again reduce the interaction to a simple energy term due to the ground state $\epsilon_0(N)=-\sqrt{u^2 N^2+p^2}$. The effective polarizer capacitance is given by the curvature of the energy profile, $C_\text{P} \sim \partial_N^2 \epsilon_0$, which is here obviously negative. Again, if the gap is not sufficient to eliminate the excited state, there will be an effective reduction of the negative capacitance, due to the positive curvature of the excited state. For a more concrete example, imagine a discrete chain of charge islands (e.g. quantum dots), which are linearly arranged away from the superconducting island. Then, the electrostatic interaction due to the charge $N$ will decay as the inverse of the distance. Suppose in addition that the onsite energy on the chain is tunable with additional gates, giving rise to a polarizer potential. Let us for simplicity take for the latter a linear dependence on the distance, $\sim j$. The resulting polarizer Hamiltonian is of the form
\begin{multline}
    \widehat{H}_\text{P} = -t \sum_{j=2}^{J} \left(\vert j\rangle\langle j - 1\vert + \vert j - 1\rangle\langle j\vert\right)\\
    + \sum_{j=1}^J \left(u \frac{N - N_0}{j} + p j\right) \vert j\rangle\langle j\vert\,,
\end{multline}
where $t$ accounts for inter-dot hopping, and the shift $N_0$ accounts for an offset charge which is not necessarily the same as $N_\text{g}$. We can estimate the energy dependence of the ground state by finding the minimum of the potential energy of the above Hamiltonian, $\epsilon_\text{min}(N)$ (i.e., neglecting for simplicity the kinetic term $\sim t$). For $N>N_0$, this amounts to an energy profile $\epsilon_\text{min}\sim \sqrt{N-N_0}$, which yet again has the sought-after negative curvature. The advantage with respect to the two-level model is that here, the energies of the lowest few excited states also exhibit a negative curvature, such that the system is less susceptible to occupations of excited states. 

Finally, another potentially successful strategy to improve the ratio between $E_\text{S}$ and $E_\text{C}$ is to replace the Josephson junction of the auxiliary transmon with a Cooper-pair cotunneling junction (CCJ, see below for details), realizing a $\sim \cos(2\widehat{\varphi}_\text{T})$ term. This decouples the even and odd Cooper-pair parity states, effectively increasing both $E_\text{CT}$ and $E_\text{S}$ by a factor of 4 while decreasing $\lambda$ by 2. This lifts the above mentioned minimum restriction on $E_\text{C} / E_\text{CT}$ for the Dirac circuit we discuss in a moment, where we need $\lambda = 1/2$. The drawback here is however that we would need a near perfect decoupling of the even and odd Cooper pair number states to avoid significant disturbance of the cosine behavior. This would require the manufacturing of CCJs without residual regular Josephson effect, which may still be challenging.

\subsection*{Dirac material}\label{sec:Dirac}

In the previous section, we have already pointed out a variety of ways to render the quasiperiodic nonlinear capacitor (QPNC) dominant. Here we want to provide an additional strategy, which, surprisingly, turns out to give rise to an interesting side effect. Namely, this will allow us to emulate a protected 2D Dirac material, importantly, by using only topologically trivial materials and circuit elements.

Consider that in the space of the superconducting phase $\widehat{\varphi}$, the QPNC term can be written in terms of phase translation operators shifting the phase by $\pm 2\pi \lambda$, while the linear capacitance corresponds to infinitesimal phase shifts analogously to an ordinary kinetic energy term but in phase space instead of position space. Due to this difference in their behavior (one acting non-locally and the other locally in phase space) we can attempt to suppress the linear capacitance without deprecating the effects of the QPNC.

This feat can be achieved as follows.
To suppress the local kinetic term of the linear capacitor we may include an additional potential energy term with minima that are well separated in phase space. If the potential barrier is high, the tunneling between the minima due to ordinary quantum phase slips by means of the $E_\text{C}$-term are then suppressed. The simplest nontrivial circuit element that fits this bill would be a Cooper-pair cotunneling junction (CCJ) [see Fig.~\ref{fig:master-figure}(c)],
described by a $\sim\cos(2\widehat{\varphi})$ energy term. Note that the $\cos(\widehat{\varphi})$-term of a regular Josephson junction does not work here since charge quantization renders the phase space $2\pi$-periodic and, hence, the single energetic minimum would also suppress any non-local phase slips. A CCJ can be realized by means of a symmetric superconducting quantum interference device (SQUID) while applying a magnetic flux of half a flux quantum $\Phi_0 / 2$ through the loop of the SQUID. This will cause the regular tunneling events to interfere destructively such that the next-higher-order terms become dominant involving the tunneling of pairs of Cooper-pairs~\cite{Doucot2002, Smith_2020}.
A separate recent realization of a CCJ is in Josephson junctions formed by twisting stacks of the optimally doped high-temperature copper oxide superconductor Bi$_2$Sr$_2$CaCu$_2$O$_{8+y}$  at an angle of $45^{\circ}$~\cite{sigrist1998,yip1995,kuboki1996,can2020hightemperature,volkov_jos,tummuru2022}. Here, the residual single Cooper-pair tunneling term precisely vanishes due to the symmetry of the $d$-wave superconductors, and the observation of the second harmonic of the Josephson relation was observed in both the Fraunhoffer pattern and fractional Shapiro steps in twist junctions near 45$^{\circ}$~\cite{frank_exp}.
Beyond these realizations, the CCJ recently enjoyed a lot of interest: on the experimental side it was used to build a modified transmon~\cite{Smith_2020}, giving rise to two ground states with the same fermion parity. On the theory side, it was proposed as a key ingredient to engineer flat-band materials~\cite{Chirolli_2021}.

While ordinary quantum tunneling between the minima of the additional $\cos(2\widehat{\varphi})$-potential will be suppressed, the nonlinear capacitor may still provide non-local tunneling processes. 
Here, this tunneling is most prominent if we set the parameters such that $\lambda = 1/2$. We also add a regular Josephson junction of energy $E_\text{J}$ in parallel, creating an asymmetry between the minima of the potential. Thus, we arrive at the Hamiltonian
\begin{multline}\label{eq:Hamiltonian_Dirac_circuit}
	\widehat{H}_\text{D} = 2 E_\text{C} \Bigl(\widehat{N} + N_\text{g}\Bigr)^2 - E_\text{S} \cos\Bigl[\pi \Bigl(\widehat{N} + N_\text{g}\Bigr) + 2\pi N_\text{gT}\Bigr]\\
	- E_\text{J} \cos\bigl(\widehat{\varphi}\bigr) - E_\text{2J} \cos\bigl[2 \bigl(\widehat{\varphi} - \delta\bigr)\bigr],
\end{multline}
where the offset charge $N_\text{g}$ is induced by the gate voltage $V_\text{g}$ connected to the island, $\delta$ is the phase difference across the circuit, and $E_\text{2J}$ is the cotunneling energy of the CCJ.

In order to create a strong confinement of the phase, we consider the cotunneling $E_\text{2J}$ to be the dominant energy scale, which---considering the aforementioned recent experimental advances~\cite{Smith_2020}---seems within reach. Hence, the lowest two energy eigenstates are localized in the minima of the cotunneling term at $\varphi_j = \delta + j\pi$ with $j \in \{0, 1\}$ and their energetic difference is much smaller than the gap to the next excited state, which, for $E_\text{2J}\gg E_\text{C}$, is achieved by assuming $E_\text{J}, E_\text{S} \ll 4\sqrt{E_\text{C}E_\text{2J}}$. In the following we make use of this energy separation of the eigenstates to reduce the size of the Hilbert space.
Namely, we are splitting the Hamiltonian into two terms $\widehat{H}_\text{D} = \widehat{H}_0 + \widehat{V}_\text{D}$ where $\widehat{V}_\text{D} = - E_\text{S} \cos\bigl[\pi \bigl(\widehat{N} + N_\text{g}\bigr) + 2\pi N_\text{gT}\bigr] - E_\text{J} \cos(\widehat{\varphi})$ is a small perturbation to $\widehat{H}_0 = 2 E_\text{C} \bigl(\widehat{N} + N_\text{g}\bigr)^2 - E_\text{2J} \cos[2 (\widehat{\varphi} - \delta)]$. We can thus first diagonalize $\widehat{H}_0$ (in principle exactly), and then take into account the effect of $\widehat{V}_\text{D}$ perturbatively, in the low-energy basis of $\widehat{H}_0$.

Since charge and phase have the same duality as position and momentum we can use Bloch's theorem to solve $\widehat{H}_0$. However, as is well-established by now~\cite{Cottet_2002}, the important difference to a regular particle in a 1D potential is the $2\pi$-periodicity constraint on the wave function $\psi(\varphi)$, in addition to the periodicity of the Hamiltonian. Since this aspect is of importance below, let us reiterate it for completeness---first, with the example of a regular transmon Hamiltonian, which is $2\pi$-periodic in $\varphi$. Any attempt to try and gauge away the offset charge $N_\text{g}$ by a unitary transformation $\widetilde{\psi}(\varphi) = \mathrm{e}^{\mathrm{i} N_\text{g} \varphi} \psi(\varphi)$ is bound to fail, as $N_\text{g}$ now simply reenters in the boundary condition of $\widetilde{\psi}(\varphi)$, ensuring the periodicity of the original wave function $\psi$. In this analogy $\widetilde{\psi}$ is a Bloch function and the offset charge $N_\text{g}$ takes the role of the quasimomentum $k$ in the solid state---except that instead of being a continuous quantum number, $N_\text{g}$ is an external parameter, selecting a particular Bloch wave function $\widetilde{\psi}_n$ from each Bloch band $n$. This gives us a discrete spectrum depending on $N_\text{g}$ instead of continuous bands. Note that in contrast to usual applications of Bloch's theorem, where the non-periodic Bloch function takes the role of the wave function, here the wave function $\psi_n$ is the periodic function that modulates the plane wave $\mathrm{e}^{\mathrm{i} N_\text{g} \varphi}$ of the Bloch function $\widetilde{\psi}_n$.

With the CCJ, the Hamiltonian is now $\pi$-periodic, but we still have the same $2\pi$-periodicity constraint on the wave function $\psi$. That is why $N_\text{g}$ now selects two states per Bloch band, $\psi_n^+$ and $\psi_n^-$ with $k = N_\text{g}$ and $N_\text{g}+1$, which are related via $\psi_n^-(\varphi) = \mathrm{e}^{\mathrm{i} \varphi} \psi_n^+(\varphi)$ and with $\psi_n^+$ being $\pi$-periodic. Thus $\psi_n^+$ ($\psi_n^-$) is (anti-)symmetric under $\pi$-translation. 
As argued above, we now focus on the two states of the lowest band, $\psi_0^\pm$. Due to large $E_\text{2J}$, we can approximate them as superpositions of two Gaussian-like wave packets each localized at a minimum $\varphi_j$. Moreover, the two states form a quasi-degenerate ground state for all values of $N_\text{g}$, up to exponentially suppressed phase slip terms, such that, when eliminating all but these two low energy states, $\widehat{H}_0\approx \text{const.}$, which can thus be set to zero (deviations from this assumption cannot destroy the effect, as we explain in a moment).
The remaining low-energy Hamiltonian is entirely due to $\widehat{V}_\text{D}$, and can be computed by projecting it onto the eigenbasis of $H_0$,
\begin{equation}\label{eq:2D_Dirac_Hamiltonian}
	\widehat{H}_{\text{D,eff}} = - E_\text{S} \cos\left(\pi N_\text{g}\right) \widehat{\sigma}_z - E_\text{J} \cos\left(\delta\right) \widehat{\sigma}_x\,.
\end{equation}
The offset charge $N_\text{gT}$ leads only to a shift in $N_\text{g}$. Apart from that, it does not affect the eigenspectrum and is therefore being omitted. Here, $\widehat{\sigma}_x$ and $\widehat{\sigma}_z$ are the Pauli matrices with $\widehat{\sigma}_z = |\psi_0^+\rangle \langle \psi_0^+| - |\psi_0^-\rangle \langle \psi_0^-|$. 

This Hamiltonian shows four Dirac points in the parameter space of $(\delta, N_\text{g})$ located at $\delta \in \{\pi/2, 3\pi/2\}$, $N_\text{g} \in \{1/2, 3/2\}$. Thus the Dirac physics is here defined in a mixed parameter space of phase and charge. This is reminiscent of an earlier work by some of the authors, simulating a Chern insulator~\cite{Herrig2022} in charge and phase space. Surprisingly, we find that, here, the interplay between the QPNC and the CCJ allows for a mechanism to protect degeneracies in a 2D parameter space, which could not have been anticipated in Ref.~\cite{Herrig2022}. To demonstrate this remarkable protection let us discuss the stability of the effect when tuning away from the ideal parameter setting.

First of all, the Dirac nature of the system is completely insensitive to any small detuning $\delta\lambda$ from the exact $\lambda = 1/2$. The only effect of such a deviation is a small reduction of the amplitude in front of the $\cos(\pi N_\text{g})$-term in Eq.~\eqref{eq:2D_Dirac_Hamiltonian} (to a value slightly below $E_\text{S}$) because it reduces the overlap of the matrix elements $\langle\psi_0^\pm\vert \widehat{V}_\text{D}\vert\psi_0^\pm\rangle$. Actually, when considering the full model, it turns out that a small detuning can even have a positive effect on the spectrum by counteracting other imperfections; see Supplementary Note 4.

Another important perturbation might come from the fact that the realization of the CCJ might not be ideal and create a residual $\delta E_\text{2J} \cos(\widehat{\varphi} - \delta)$ term. In Ref.~\cite{Smith_2020}, this term is important, as it could lift the degeneracy of the two same-parity ground states of $H_0$. In our system however, yet again, small perturbations of this kind only shift the Dirac points but cannot gap the spectrum. The perturbed Hamiltonian is of the form
\begin{equation}
	\widehat{H}_{\text{D,eff}} = - E_\text{S} \cos\left(\pi N_\text{g}\right) \widehat{\sigma}_z - \left[E_\text{J} \cos\left(\delta\right) + \delta E_\text{2J}\right] \widehat{\sigma}_x\,,
\end{equation}
where the Dirac points are shifted to $\delta_\text{c}$ and $2\pi - \delta_\text{c}$, with $\delta_\text{c} = \arccos(-\delta E_\text{2J} / E_\text{J})$. As long as the residual term is smaller than the CCJ term, $\delta E_\text{2J} < E_\text{J}$, the spectrum is not gapped, and for a small $\delta E_\text{2J} \ll E_\text{J}$ we find a linear shift $\delta_\text{c} = \pi / 2 + \delta E_\text{2J} / E_\text{J}$. 
Finally, we note that the position of the Dirac points can also minimally shift when deviating from the ideal regime $E_\text{2J}\gg E_\text{C}$, which slightly lift the degeneracy of the two low-energy basis states of $H_0$ (thus adding a small contribution in $\widehat{\sigma}_z$). 

Now we want to study how one can probe the presence of the Dirac points in a straightforward fashion. We show that in this particular case, the Dirac physics is responsible for a change in the asymptotic behavior of the classical current-noise signal. 
We consider a time-dependent driving of the offset charge $N_\text{g}$. Consequently, there will be a geometric component of the current response which can be cast into the form of a Berry curvature~\cite{Riwar2016,Herrig2022} in the $(\delta, N_\text{g})$ parameter space. Importantly, the Dirac material has the special property that the Berry curvature vanishes exactly for each point $(\delta, N_\text{g})$, except for the Dirac points. Away from these points the Berry curvature is nonzero only when the Dirac cones are gapped. We therefore need to find a quantity capable of measuring the presence or absence of the Berry curvature in a local fashion. As we show in the following, this can be accomplished by measuring the classical current noise spectrum $S\left(\omega\right) = \int_{-\infty}^{\infty} d\tau \exp\left(\mathrm{i} \omega \tau\right) \overline{I(t) I(t + \tau)}$ as a response to fluctuations in $N_\text{g}$.

\begin{figure}
	\centering
	\includegraphics[width=.8\linewidth]{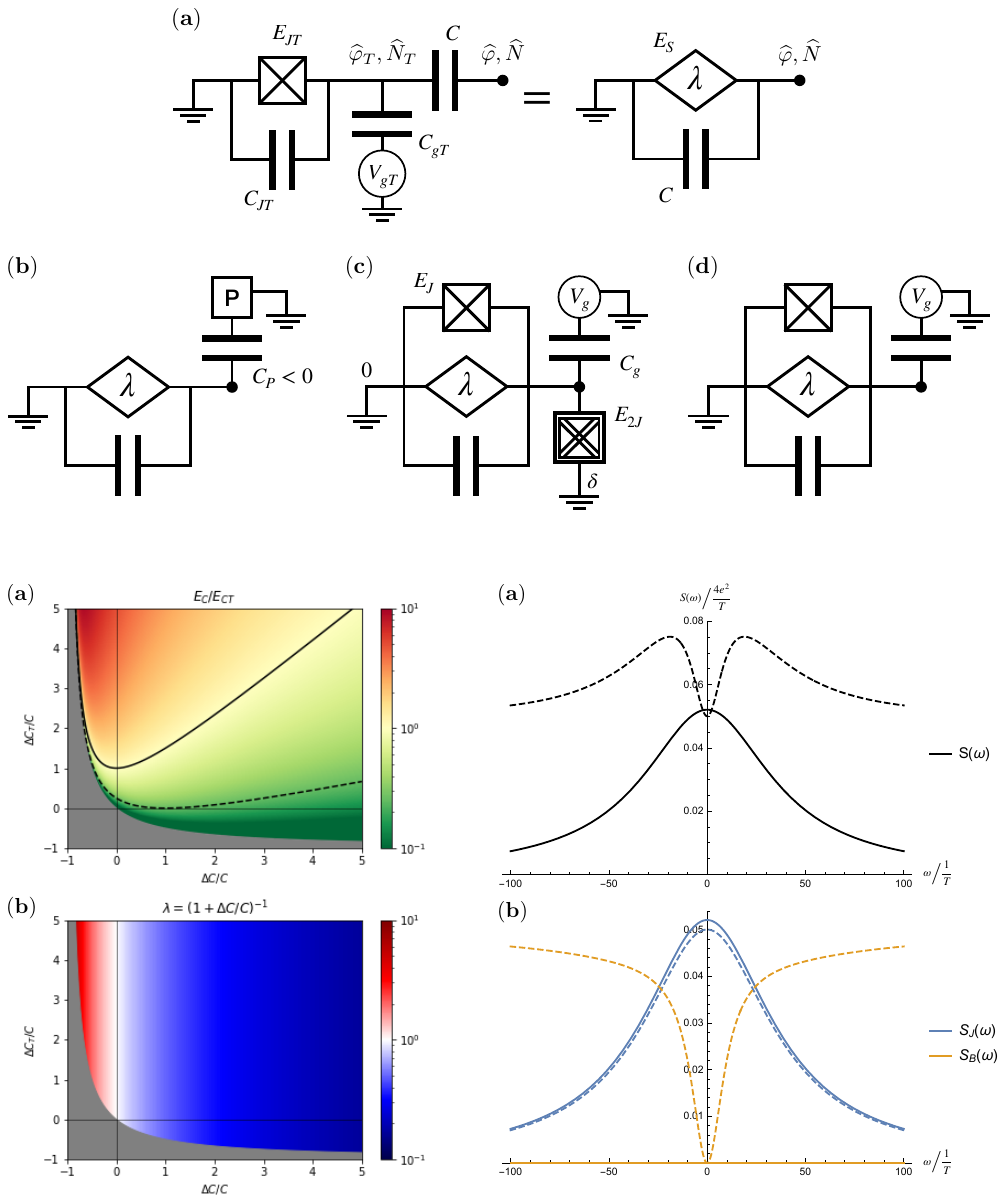}
	
	\caption{\textbf{Classical current noise spectrum.} (a) The full noise spectrum $S(\omega)$ can be split into (b) two contributions, $S_\text{J}(\omega)$ from the Josephson current and $S_\mathcal{B}(\omega)$ from the Berry curvature, corresponding to the adiabatic approximation up to the first correction. Depicted are two scenarios---the solid lines show what we expect in our system while for the dashed lines, we introduced a small artificial gap $\Delta$ of the Dirac points in form of an $\widehat{\sigma}_y$-term. As clearly shown here, in contrast to the former, the latter system produces a finite Berry curvature leading to a finite asymptotic value of high frequencies $\omega$. We chose a gap of $\Delta = 1.06 T^{-1}$ with $E_\text{S} = E_\text{J} = 10 T^{-1}$ and $\delta = \pi / 4$ such that $\beta = S(\infty) / S(0) \approx 1$; see Supplementary Note 5.
	}
	\label{fig:noise-spectrum}
\end{figure}

We describe the offset charge noise, $N_\text{g} = N_\text{g}^{(0)} + \xi(t)$, by white noise such that $\overline{\partial_t \xi(t)} = 0$ and $\overline{\partial_t \xi(t) \partial_t \xi(t')} = 2 \delta (t-t') / T$. Here, $\overline{\bullet}$ denotes the ensemble average and $T$ characterizes the magnitude of the noise correlations and has to be determined experimentally~\cite{Serniak2018,Serniak2019,Christensen2019}.
Assuming slow noise (that is, $\partial_t \xi(t)$ being always smaller than the local gap size) 
and starting in the ground state $\left|0\right\rangle$, we use an adiabatic approximation of the wave function~\cite{Thouless1983} to calculate the expectation value of the current $\widehat{I} = 2e\, \partial_\delta \widehat{H}$ up to first order in $\dot{N}_\text{g}$
\begin{equation}
	I \left(t\right) = I_0 \left(t\right) + I_1 \left(t\right), \label{eq:current-adiabatic}
\end{equation}
with the Josephson current $I_0 (t) = \langle 0| \widehat{I} |0\rangle = 2e\, \partial_\delta \epsilon_n$ and the linear correction term $I_1 (t) = -2e \dot{N}_\text{g} \mathcal{B}_0 [\delta, N_\text{g} (t)]$. Here, $\epsilon_n$ are the eigenvalues of the Hamiltonian with corresponding eigenstates $|n\rangle$ and the Berry curvature is given by $\mathcal{B}_n (\delta, N_\text{g}) = -2 \operatorname{Im} \bigl\langle \partial_\delta n \big| \partial_{N_\text{g}} n \bigr\rangle$. 
The periodicity of the current in $N_\text{g}$ can be exploited by casting the contributions into a discrete Fourier series; for details see Supplementary Note 5. Ultimately, we find two contributions to the current noise spectrum, one from the Josephson current (with subscript J) and a Berry curvature contribution (with subscript $\mathcal{B}$)
\begin{equation}\label{eq:current-noise-spectrum_main-text}
	S\left(\omega\right) = S_\text{J}\left(\omega\right) + S_{\mathcal{B}}\left(\omega\right).
\end{equation}
The respective asymptotic behavior (the power law) can be computed by means of the respective dominant Fourier component. For the Josephson current contribution, we get
\begin{equation}
	S_\text{J} \sim \omega^{-2}\,,
\end{equation}
whereas the Berry curvature contribution yields
\begin{equation}
	S_{\mathcal{B}} \sim \text{const.}\,,
\end{equation}
where the respective constants of proportionality are quadratic in the respective dominant Fourier coefficient. Note that we exclude from these considerations the zeroth Fourier coefficients corresponding to the time-averaged current contributions---they naturally do not contribute to the noise spectrum. Moreover, in the proximity of the degeneracies one might has to consider multiple Fourier coefficients for a quantitative analysis; see Supplementary Note 5.
Now, depending on whether the Berry curvature is zero or nonzero, this changes the asymptotic behavior of the classical noise for large $\omega$.
The total spectrum $S$ and its two contributions $S_\text{J}$ and $S_{\mathcal{B}}$ are shown for the Dirac system and a more general gapped system in Fig.~\ref{fig:noise-spectrum}. If the Berry curvature is zero, the noise decays as $\omega^{-2}$. If it is nonzero, it goes to a constant value. 
One way to realize the latter case in an experiment is to consider a decreased $E_\text{2J}$ which introduces a gap into the spectrum.
To conclude, we can probe whether or not the system is gapped by measuring the asymptotic value of the classical current noise spectrum.

\begin{figure}
	\centering
	\includegraphics[width=\linewidth]{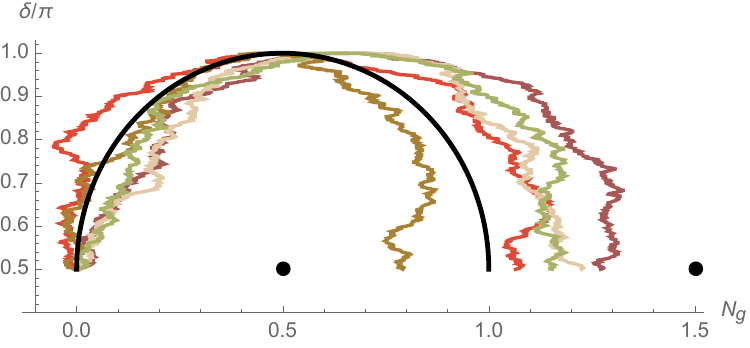}
	
	\caption{\textbf{Noisy semicircle path around a Dirac cone.} Depicted in colors are five exemplary paths with possible realizations of the offset-charge noise. The ideal path (see black line) halfway encircles exactly one Dirac point. However, as long as the actual path starts and ends at the value of $\delta$ where the degeneracy is located (here $\delta=1/2$), while passing exactly one degeneracy, the Berry phase picked up by the two states differs by $\pi$, independent of the exact shape of the parameter curve. Hence, as long as the driving occurs sufficiently faster than the offset charge fluctuations, or the offset charge itself can be sufficiently controlled to avoid the Dirac points, the measurement of the Berry phase can be achieved.
	}
	\label{fig:noisy-path}
\end{figure}

For the sake of completeness, let us briefly sketch a different approach to measure the presence of the Dirac points, via measuring the Berry phase. Namely, while the Berry curvature $\mathcal{B}$ away from the degeneracies is exactly zero, taking an adiabatic path around a Dirac point, where $\mathcal{B}$ is singular, allows us nonetheless to pick up a Berry phase. However, since a full circle will give us a phase of $\pi$ for both states, we cannot measure any phase difference. Therefore, one would have to perform a semicircle around the degeneracy in $(\delta, N_\text{g})$-space, where the two states gather a phase of $\pm \pi/2$ giving a difference of $\pi$. We do not discuss this possibility in more detail due to experimental challenges, in particular the large amplitude off-set charge noise which impede a precise control of $N_\text{g}$~\cite{Serniak2018,Serniak2019,Christensen2019}. As a matter of fact, the very same offset-charge noise was a necessary ingredient for the previously described current noise probe. Nonetheless, via active measurement of background charges and backaction control, it is perceivable that a sufficient time-dependent control of $N_\text{g}$ could potentially be achieved. Then, to realistically perform this measurement, one should start and end the evolution at a value of $\delta$ at a degeneracy. Thus, we do not care about the exact value of $N_\text{g}$ as long as the aforementioned feedback control allows to sufficiently fix the value of $N_\text{g}$ to  stay away from the degeneracies. One could then tune $\delta$ and $N_\text{g}$ around a single cone; see Fig.~\ref{fig:noisy-path}.
Since the Berry phase depends only on the angle around the cone and not the exact shape of the parameter curve, the topological feature could be observed.

\subsection*{Charge localization in the Aubry-Andr\'e model}\label{sec:aubry-andre}

Previously, we interpreted the Josephson terms as potentials, identifying the phase as position space (with torus topology), while capacitive terms were taking the role of kinetic energies. In this section we invert this identification, interpreting the capacitive terms as potentials on an infinite charge lattice while the Josephson terms introduce a hopping on this lattice.

\begin{figure*}
	\centering
	\includegraphics[width=.97\linewidth]{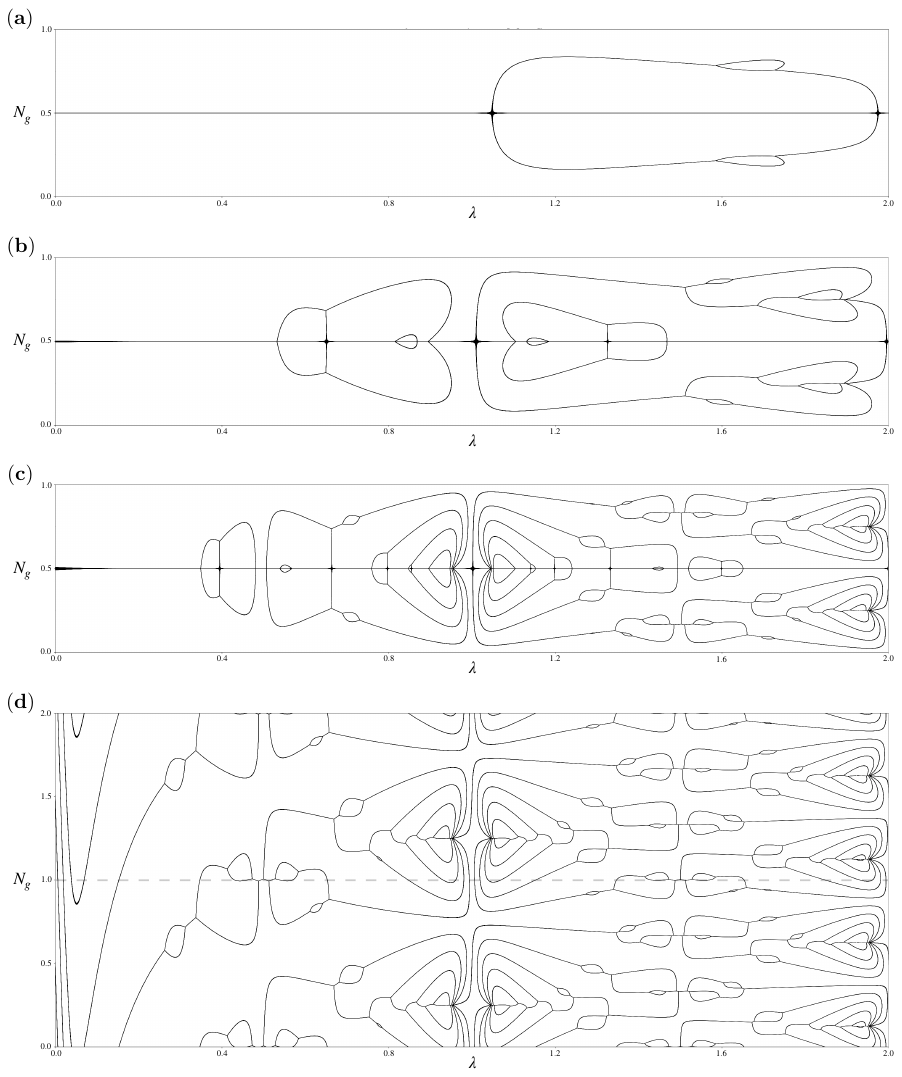}
	
	\caption{\textbf{Charge fluctuation anomalies related to charge localization in the Aubry-Andr\'e model.} For all white areas, the charge is localized with negligible charge fluctuations. The black lines represent the values for the parameter pair $(\lambda, N_\text{g})$ for which the ground state is degenerate and, thus, the quantum fluctuations of the charge are finite, ultimately leading to delocalized states. This breaking of the charge localization is shown for $N_\text{gT} = 0$ and the energy ratios (a) $E_\text{S} = E_\text{C}$, (b) $E_\text{S} = 5 E_\text{C}$, and (c) $E_\text{S} = 25 E_\text{C}$, as well as for (d) $N_\text{gT} = 1/4$ and $E_\text{S} = 25 E_\text{C}$. We see that the complexity of the pattern increases for increasing $E_\text{S} / E_\text{C}$. Moreover, $N_\text{gT} \neq 0$ breaks the mirror symmetry along $N_\text{g}$ but not the periodicity.
	}
	\label{fig:charge-anomalies}
\end{figure*}

Due to the charge operator being quantized, the nonlinear capacitor is only truly periodic if $\lambda$ is a rational number. For irrational values, the associated energy term becomes quasiperiodic and by that realizes a quasiperiodic lattice in charge space. 
This opens the possibility to study the interplay of this lattice with kinetic energy implemented by a Josephson junction with energy $E_\text{J}$ in parallel to the capacitor [see~Fig.~\ref{fig:master-figure}(d)], giving us the Hamiltonian
\begin{equation}\label{eq:AA}
    \widehat{H}_{\text{AA}} = \widehat{T} + \widehat{V}\,,
\end{equation}
with the on-site energy of the quasiperiodic lattice
\begin{multline}
	\widehat{V} = - E_\text{S} \cos\left(2\pi \left[\lambda \left(\widehat{N} + N_\text{g}\right) + N_\text{gT}\right]\right)\\
	+ 2E_\text{C} \left(\widehat{N} + N_\text{g}\right)^2,
\end{multline}
and the Josephson energy $\widehat{T} = - E_\text{J} \cos(\widehat{\varphi})$. We can rewrite the latter in the charge eigenbasis $|N \rangle$ as a hopping term on the lattice
\begin{equation}
    \widehat{T} = - \frac{E_\text{J}}{2} \sum_N \left|N \right\rangle \left\langle N - 1\right| + \text{h.c.}\,,
\end{equation}
due to the quantization condition between charge and phase, $\bigl[\widehat{\varphi}, \widehat{N}\bigr] = \mathrm{i}$.

The model, Eq.~\eqref{eq:AA}, at $E_\text{C} = 0$ corresponds to the Aubry-Andr\'e (AA) model and $E_\text{C} >0$ adds a quadratic trapping potential, which is qualitatively similar to an ultra-cold atom realization of the AA model~\cite{RoatiInguscio2008,Luschen:2018aa,Wang:2022aa}. 
In the absence of the potential ($E_\text{C}=0$) the AA model is exactly solvable with a delocalized phase for $E_\text{S}<E_\text{J}$, an Anderson localized phase for $E_\text{S}>E_\text{J}$, and all the eigenstates are critical at the (self dual) critical point $E_\text{S}=E_\text{J}$.
It is has been shown that for $E_\text{C} \ll E_\text{S}$, this trapping potential induces finite size effects, in particular a rounding out of the transition, but the wave functions for $E_\text{J} \gg E_\text{S}$ ($E_\text{J} \ll E_\text{S}$) qualitatively retain their delocalized (Anderson localized) characteristics~\cite{Modugno2009}. It is worthwhile emphasizing that Anderson localization only occurs for irrational $\lambda$: Otherwise,  the $E_\text{S}$-term merely introduces a `superlattice width', a rescaled (larger) lattice spacing, which in turn implies delocalized Bloch waves living in an appropriately rescaled (smaller) Brillouin zone.
On the other hand, of course, once $E_\text{C} \gg E_\text{S}$ the AA physics is lost. However, as illustrated in Fig.~\ref{fig:negative-capacitances} above, with the help of negative capacitances (e.g. realized with the polarizer concept outlined above), we can also tune the relative energy scales to minimize the impact of the trap.

In this system, Anderson localization will manifest itself in the form of charge localization. This effect could be probed in a number of ways, e.g. by initializing the circuit in a given charge state, and subsequently measuring the dispersion (or lack thereof) in charge space as a function of time. We here choose to compute a related, but even simpler quantity. Namely, we assume that the system is prepared in the ground state, and compute the quantum charge fluctuations associated with it. Here, charge localization manifests very simply as
\begin{equation}
    \Delta N \equiv \langle \widehat{N}^2 \rangle - \langle \widehat{N} \rangle ^2 = 0\,. \label{eq:DeltaN}
\end{equation}
This effect is only broken if $\widehat{V}$ becomes degenerate in its ground state, which happens for specific parameter values of $\lambda$, $N_\text{g}$, and $N_\text{gT}$. Here, the charge fluctuations of the ground state immediately spike to a finite value. This gives rise to an interesting pattern of suppressed charge fluctuations, which are lifted along characteristic, rather complex contours, which we refer to as charge fluctuation anomalies. As an illustration of the regimes of avoided groundstate degeneracies, and hence strong localization physics, see Fig.~\ref{fig:charge-anomalies}.

\begin{figure}
    \centering
    \includegraphics[width=.85\linewidth]{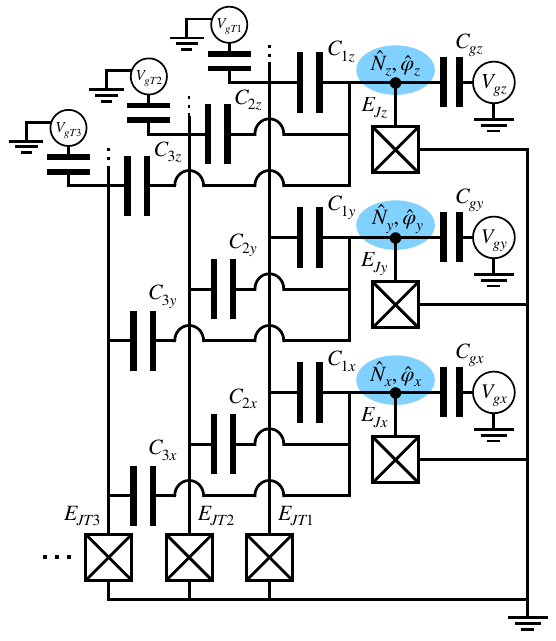}
    
    \caption{\textbf{Circuit implementing the Aubry-Andr\'e model in multiple dimensions.} Shown is the 3D version but the circuit can be generalized to arbitrary higher dimensions by continuing the structure. In $d$ dimensions we need $d$ auxiliary transmons (with a Josephson junction of energy $E_{\text{JT},j}$) each capacitively connected (via capacitance $C_{j\mu}$, $\mu \in \{x, y, \ldots\}$) to $d$ islands (marked blue) with the corresponding charges ($\widehat{N}_x$, $\widehat{N}_y$,~...) and the phases ($\widehat{\varphi}_x$, $\widehat{\varphi}_y$,~...) forming a $d$-dimensional charge and phase vector, $\vctr{\widehat{N}}$ and $\vctr{\widehat{\varphi}}$, respectively. Each island is subject to an offset charge $N_{\text{g},\mu} = C_{\text{g},\mu} V_{\text{g},\mu} / 2e$ and connected to ground via a Josephson junction of energy $E_{\text{J},\mu}$, both of which also form a vector $\vctr{N}_\text{g}$ and $\vctr{E}_\text{J}$.}
    \label{fig:multidim_Aubry-Andre}
\end{figure}

As mentioned in the introduction, higher dimensional analogs of the AA model are much less understood. In particular, the nature of quasiperiodic localization in higher dimensions and its relation to the more conventional, disorder driven localization is an interesting, timely and largely open question. Therefore, it is quite important to have experimental settings where these ideas can be tested and explored.
A crucial advantage of the platform offered by superconducting circuits is the possibility to realize a model of in principle \textit{arbitrary} dimension~\cite{DevakulHuse2017}; the 3D version of the circuit is depicted in Fig.~\ref{fig:multidim_Aubry-Andre} but it can be generalized to dimension $d$ by continuation indicated with the dots. 
In particular, we show that one can generalize the quasiperiodic nonlinear capacitor to a multidimensional version, coupling several islands in a single cosine term. Each of the $d$ auxiliary transmons is connected to $d$ islands, whose charge and phase operators are condensed in a charge vector $\vctr{\widehat{N}} \equiv \bigl(\widehat{N}_x, \widehat{N}_y, \ldots\bigr)^\intercal$ and a phase vector $\vctr{\widehat{\varphi}} \equiv \bigl(\widehat{\varphi}_x, \widehat{\varphi}_y, \ldots\bigr)^\intercal$, as well as corresponding offset charges $\vctr{N}_\text{g}$ and Josephson energies $\vctr{E}_\text{J}$ connected to these islands. This realizes the effective Hamiltonian
\begin{equation}\label{eq:multidim-AA}
    \widehat{H}_{\text{AA}}^d = - \vctr{E}_\text{J} \cdot \cos\left(\vctr{\widehat{\varphi}}\right) + \sum_{j=1}^d \widehat{V}_j^d + \widehat{V}_\text{C}^d\,,
\end{equation}
where we define the cosine of a vector component-wise $\Bigl[\cos\Bigl(\vctr{\widehat{\varphi}}\Bigr)\Bigr]_{\mu} := \cos\Bigl(\widehat{\varphi}_{\mu}\Bigr)$.
The on-site energy contribution due to the $j$-th transmon is given by
\begin{equation}
	\widehat{V}_j^d = - E_{\text{S},j} \cos\left[2\pi \vctr{\lambda}_{j} \cdot \left(\vctr{\widehat{N}} + \vctr{N}_\text{g}\right) + 2\pi N_{\text{gT},j}\right],
\end{equation}
and the linear-capacitive contribution
\begin{equation}
	\widehat{V}_\text{C}^d = 2\left(\vctr{\widehat{N}} + \vctr{N}_\text{g}\right)^\intercal \mtrx{E}_\text{C} \left(\vctr{\widehat{N}} + \vctr{N}_\text{g}\right),
\end{equation}
where $\mtrx{E}_\text{C}$ is a matrix representing the entire capacitive network of the circuit (also containing capacitive cross-couplings between the islands). We expect the circuit to include also cross-coupling between the transmons which, according to a first analysis, appear to be suppressed in the transmon regime of dominant $E_{\text{JT},j}$. The parameters $\vctr{\lambda}_{j}$, $\mtrx{E}_\text{C}$, and $E_{\text{S},j}$ can all be calculated with the capacities of the circuit, analogously to above; see Supplementary Note 1 for a derivation. By tuning those capacities (shown in Fig.~\ref{fig:multidim_Aubry-Andre}) we can tune the components of all quasiperiodicity parameters
\begin{equation}
    \lambda_{j,\mu} = \frac{C_{j\mu}}{\sum_{j^\prime} \left(C_{j^\prime\mu} + C_{j^\prime,\text{tot}}\right)}\,,\quad \mu \in \{x, y, \ldots\}
\end{equation}
independent of each other, and thus, by choosing linearly independent vectors $\vctr{\lambda}_{j}$ with only irrational and nonzero elements, we realize a non-separable quasiperiodic model. Moreover, by switching a transmon Josephson junction $E_{\text{JT},j}$ with one of the capacities coupling this transmon to the islands, we can tune each $\vctr{\lambda}_{j}$ to have a single negative component, as already mentioned in the discussion following Eq.~\eqref{eq:lambda}. This allows for orthogonal vectors realizing a self-dual model according to \cite{DevakulHuse2017}.
As above we can use partial negative capacitances to improve the ratio of the parameters, rendering $\mtrx{E}_\text{C}$ negligible. Negative cross capacitances are a further interesting concept to explore in future works, as they would allow for the mitigation of unwanted cross talk, or (as already mentioned) present an alternative strategy to flip the sign of a specific component in the $\vctr{\lambda}_{j}$ vector.

\subsection*{Discussion}\label{sec:outlook}

In summary, we have introduced a circuit element, the quasiperiodic non-linear capacitor, which can be realized with an appropriately designed combination of capacitors and superconductor-insulator-superconductor junctions. We have discussed potential practical obstructions and mitigations thereof and outlined two applications: A Dirac material emulator in combined charge-flux space as well as an emulator of Anderson localization within the 1D and higher dimensional Aubry-Andr\'e paradigm.

We conclude with a perspective on the quantum emulation of twisttronics and the possibilities to explore and exploit synergies between the quasiperiodic circuit element and circuit emulation of topological band structures~\cite{Riwar2016}. Specifically, it has been demonstrated ~\cite{FuPixley2020,GonzalezCirac2019,ChouPixley2020,LeePixley2022} that the magic-angle effect in twisted bilayer graphene (i.e. the massive flattening of Dirac bands to virtually vanishing kinetic energy) can be emulated in a variety of systems even without the need of physically twisted atomic lattices. Moreover, near the magic angle, quasiperiodicity was theoretically found to be relevant and to lead to an eigenstate quantum phase transition analogous to Anderson localization-delocalization transitions, but in momentum rather than coordinate space. While the small absolute value of the magic twist-angle in graphene leads to an extremely narrow parameter regime governed by quasiperiodic effects, this restriction is generally absent in quantum emulators. It will thus be an interesting topic for the future to study experimental signatures of the interplay of quasiperiodicity and Dirac physics in quantum circuit analogs of magic angle twisted bilayer and multilayer graphene.

On a final note, as pointed out in the main text already, we currently consider the transport degrees of freedom to simulate the position of a single particle moving in a quasiperiodic potential. Given the importance of the interplay between quasiperiodicity and strong correlations, it is therefore a further important question, whether many-body interaction effects can be taken into account in a meaningful way in the here proposed platform. We expect that such an endeavor would in principle work similar in spirit to the generalization to higher dimensions, see Fig.~\ref{fig:multidim_Aubry-Andre}. For instance, instead of interpreting the wave function of the circuit to simulate a particle in an $n$-dimensional space, it could represent $n$ particles in a 1-dimensional space (or any other commensurable combination of spatial dimensions and particle numbers). The crucial challenge that needs to be overcome for this idea to work, is that the circuit wave function (if interpreted as representing multiple particles) needs to satisfy fermionic, bosonic, or potentially even anyonic commutation relations, which is not naturally guaranteed (the circuit wave function generically is not required to satisfy such symmetry constraints). It will have to be investigated in future works if and how such constraints can be included in the simulation. Such an approach will in all likelyhood not be able to feasibly mimic a system in the thermodynamic limit~\cite{Lee_1985,Belitz_1994}, but we expect that dilute systems~\cite{Weinmann_1997,Bourgain_2019} could be tackled.

\section*{DATA AVAILABILITY}
The data that support the findings of this study are available from the corresponding author upon reasonable request.

\begin{acknowledgments}
	We acknowledge fruitful discussions with P.\ Bushev. This work has been funded by the German Federal Ministry of Education and Research within the funding program Photonic Research Germany under the contract number 13N14891.
	J.H.P.\ is partially supported by the Air Force Office of Scientific Research under Grant No.~FA9550-20-1-0136 and the Alfred P.\ Sloan Foundation through a Sloan Research Fellowship. The Flatiron Institute is a
division of the Simons Foundation.
\end{acknowledgments}

\section*{COMPETING INTERESTS}
The authors declare no competing financial or non-financial interests.

\section*{AUTHOR CONTRIBUTIONS}
EJK and RPR conceived of and lead the research project. TH performed all calculations under the guidance of JHP, EJK and RPR. All authors contributed to the writing of the text and reviewed the final manuscript.

\bibliographystyle{naturemag}

\bibliography{paper2}

\clearpage

\setcounter{equation}{0}
\setcounter{figure}{0}
\setcounter{section}{0}
\setcounter{table}{0}
\setcounter{page}{1}
\makeatletter
\renewcommand{\theequation}{S\arabic{equation}}
\renewcommand{\thefigure}{S\arabic{figure}}
\renewcommand{\thepage}{S\arabic{page}}

\begin{widetext}
\begin{center}
{\large
\textbf{Supplementary Information on ``Quasiperiodic circuit quantum electrodynamics''}}\\
\vspace{9pt}
T.~Herrig, J.~H.~Pixley, E.~J.~K\"onig, and R.-P.~Riwar
\end{center}
\end{widetext}

\section*{Supplementary Note 1}
\subsection*{Derivation of the Hamiltonians via the Lagrangian approach}
\subsubsection*{Quasiperiodic nonlinear capacitor}
The Hamiltonian in Eq.~(3) of the main text characterizes the circuit in Fig.~1(a) of the main text displaying the quasiperiodic nonlinear capacitor (QPNC). To find this Hamiltonian, we first use the standard rules to write down the Lagrangian
\begin{multline}
    \mathcal{L}	= \frac{1}{2} \frac{C\left(\dot{\varphi}_\text{T} - \dot{\varphi}\right)^{2}}{\left(2e\right)^{2}} + \frac{1}{2} \frac{C_\text{gT} \left(\dot{\varphi}_\text{T} - 2e V_\text{gT}\right)^{2}}{\left(2e\right)^{2}} + \frac{1}{2} \frac{C_\text{JT} \dot{\varphi}_\text{T}^{2}}{\left(2e\right)^{2}}\\
	+ E_\text{JT} \cos\left(\varphi_\text{T}\right) + \frac{1}{2} \sum_{\alpha} \frac{C_{\alpha} \left(\dot{\varphi} - \dot{\varphi}_{\alpha}\right)^{2}}{\left(2e\right)^{2}} - V(\varphi),
\end{multline}
where the last two terms correspond to additional elements connected to the right island (with phase $\varphi$) not explicitly shown in the figure. We include these here since the circuit is meant to be a part of a larger circuit and to generalize the derivation and demonstrate what influence such elements have on the parameters of the final Hamiltonian. Here, $V(\varphi)$ encapsulates any inductive contributions while the $C_{\alpha}$ denote the capacities of any linear capacitance to an island or contact with superconducting phase $\varphi_{\alpha}$. We neglect any dynamics of these phases -- that is, beyond external control like an applied voltage to ground -- which also allows us to truncate any energy terms that do not depend on $\varphi$ or $\varphi_\text{T}$. Treating the phases $\varphi_{\alpha}$ as degrees of freedom within the Lagrange formalism would result in a much more complex calculation shifting the focus away from the QPNC. Part of this complexity is the appearance of capacitive cross-couplings between the auxiliary transmon island (with phase $\varphi_\text{T}$) and the other islands whose strengths would have to be considered depending on the parameter regime. Here, we want to concentrate on the core behavior of the QPNC in circuits with few degrees of freedom. As an example of a more complex circuit we consider the multidimensional Aubry-Andr\'e circuit (see Fig.~6 of the main text); a derivation of the Hamiltonian for $d=2$ can be found below.

Now we continue with the Euler-Lagrange equations $N = \partial\mathcal{L} / \partial\dot{\varphi}$ and $N_\text{T} = \partial\mathcal{L} / \partial\dot{\varphi_\text{T}}$ to define the canonically conjugate charges,
\begin{align}
	N =& \frac{C \left(\dot{\varphi} - \dot{\varphi}_\text{T}\right)}{\left(2e\right)^{2}} + \sum_{\alpha} \frac{C_{\alpha} \left(\dot{\varphi} - \dot{\varphi}_{\alpha}\right)}{\left(2e\right)^{2}}\,,\label{eq:Euler-Lagrange-1d-QPNC-1}\\
	N_\text{T} =& \frac{C \left(\dot{\varphi}_\text{T} - \dot{\varphi}\right)}{\left(2e\right)^{2}} + \frac{C_\text{gT} \left(\dot{\varphi}_\text{T} - 2eV_\text{gT}\right)}{\left(2e\right)^{2}} + \frac{C_\text{JT} \dot{\varphi}_\text{T}}{\left(2e\right)^{2}}\,,\label{eq:Euler-Lagrange-1d-QPNC-2}
\end{align}
and define the offset charges $N_\text{g}\equiv \sum_{\alpha} C_{\alpha} \dot{\varphi}_{\alpha} / \left(2e\right)^{2}$ and $N_\text{gT} \equiv C_\text{gT} V_\text{gT} / 2e$. Therewith the Hamiltonian can be derived with a Legendre transformation $H = N \dot{\varphi} + N_\text{T} \dot{\varphi}_\text{T} - \mathcal{L}$. The kinetic part (i.e. the capacitive terms) of the Hamiltonian can also be written as $T = \vctr{Q}^{T} \mtrx{C}^{-1} \vctr{Q} / 2$ with the charge vector $\vctr{Q} = 2e \left(N + N_\text{g}, N_\text{T} + N_\text{gT}\right)^\intercal$ and the capacitance matrix $\mtrx{C}$. The latter can be constructed from the Euler-Lagrange equations such that $\vctr{Q} = \mtrx{C} \vctr{V}$ with the voltage vector $\vctr{V} = \left(\dot{\varphi}, \dot{\varphi}_\text{T}\right)^\intercal / 2e$. With Supplementary Equations~\eqref{eq:Euler-Lagrange-1d-QPNC-1} and \eqref{eq:Euler-Lagrange-1d-QPNC-2} we identify
\begin{equation}
	\mtrx{C} =
	\begin{pmatrix}
		C_{\text{tot}} & - C\\
		- C & C_{\text{T,tot}}
	\end{pmatrix},
\end{equation}
introducing the total capacities of the islands $C_{\text{tot}} \equiv C + \sum_{\alpha} C_{\alpha}$ and $C_{\text{T,tot}} \equiv C + C_\text{gT} + C_\text{JT}$, and the inverse
\begin{equation}
	\mtrx{C}^{-1}=\frac{1}{C_{\text{tot}}C_{\text{T,tot}}-C^{2}}\begin{pmatrix}C_{\text{T,tot}} & C\\
		C & C_{\text{tot}}
	\end{pmatrix}.
\end{equation}
This results in the Hamiltonian
\begin{multline}\label{eq:general_Hamiltonian}
    H = 2 E_\text{CT} \left[N_\text{T} + N_\text{gT} + \lambda \left(N + N_\text{g}\right)\right]^{2}\\
    + 2 E_\text{C} \left(N + N_\text{g}\right)^{2} - E_\text{JT} \cos\left(\varphi_\text{T}\right) + V(\varphi),
\end{multline}
with the parameter definitions $E_\text{CT} = e^2 C_\text{tot} / (C_\text{tot} C_{\text{T,tot}} - C^2)$, $E_\text{C} = e^2 / C_\text{tot}$ and $\lambda = C / C_\text{tot}$, analogously to the main text. Comparing this Hamiltonian (after quantizing) with Eq.~(3) of the main text we find full agreement when removing the additional terms by setting $V(\varphi) = 0$ and $\sum_{\alpha} C_\alpha = 0$, resulting in $N_\text{g} = 0$ and $C_{\text{tot}} = C$. However, by including these terms we find out how every capacitive element connected to the right island contributes via $C_{\text{tot}}$ to the value of $\lambda$ and tunes it away from the trivial value $1$.

Now the two circuits in Figs.~1(c, d) of the main text are examples for complete circuits involving the QPNC that can be described by the general Hamiltonian in Supplementary Equation~\eqref{eq:general_Hamiltonian}. For the Aubry-Andr\'e circuit (Fig.~1(d) of the main text) we include a Josephson junction to ground (giving us the inductive term $V(\varphi) = -E_\text{J} \cos\left(\varphi\right)$) and a capacitor connected to a gate voltage to ground (setting the total capacity $C_{\text{tot}} = C + C_\text{g}$ and the offset charge $N_\text{g} = C_\text{g} V_\text{g} / 2e$).
The same elements are included in the Dirac circuit (Fig.~1(c) of the main text), however, an additional cotunneling junction connects to ground with a phase of $\phi_\alpha = \delta$. Here the full inductive potential reads $V(\varphi) = -E_\text{J} \cos\left(\varphi\right) - E_\text{2J} \cos\left[2 \left({\varphi} - \delta\right)\right]$.
Note that the Josephson junctions and the cotunneling junction are also associated with capacities which are added to $C_{\text{tot}}$, too, but for simplicity we do not include them explicitly. After quantizing the Hamiltonians we just replace the terms containing $\widehat{N}_\text{T}$ and $\widehat{\varphi}_\text{T}$ by the low-energy effective treatment $\sim E_\text{S}$ introduced and discussed in and around Eq.~(5) of the main text to find the Hamiltonians shown in Eqs.~(9) and (16) of the main text.

\subsubsection*{2D Aubry-Andr\'e circuit}
Here we derive the Hamiltonian given in Eq.~(20) of the main text for the $d=2$ dimensional Aubry-Andr\'e circuit; see Fig.~6 of the main text. The Lagrangian reads
\begin{multline}
	\mathcal{L}_\text{AA}^{(2)} = \frac{1}{2} \sum_{\mu\in\left\{ x,y\right\} } \frac{C_{\text{g},\mu} \left(\dot{\varphi}_{\mu} - 2e V_{\text{g},\mu}\right)^{2}}{\left(2e\right)^{2}}\\
	+ \frac{1}{2} \sum_{j,\mu} \frac{C_{j\mu} \left(\dot{\varphi}_{\text{T},j} - \dot{\varphi}_{\mu}\right)^{2}}{\left(2e\right)^{2}} + \frac{1}{2} \sum_{j} \frac{C_{\text{gT},j}\left(\dot{\varphi}_{\text{T},j} - 2e V_{\text{gT},j}\right)^{2}}{\left(2e\right)^{2}}\\
	+ \sum_{j} E_{\text{JT},j} \cos\left(\varphi_{\text{T},j}\right) + \sum_{\mu} E_{\text{J},\mu} \cos\left(\varphi_{\mu}\right),
\end{multline}
where each index takes two values $j\in\left\{ 1,2\right\}$ and $\mu\in\left\{ x,y\right\}$. Thus we can derive the following four Euler-Lagrange equations
\begin{align}
	N_{\mu} &= \sum_{j} \frac{C_{i\alpha} \left(\dot{\varphi}_{\mu} - \dot{\varphi}_{\text{T},j}\right)}{\left(2e\right)^{2}} + \frac{C_{\text{g},\mu} \left(\dot{\varphi}_{\mu} - 2e V_{\text{g},\mu}\right)}{\left(2e\right)^{2}}\,,\label{eq:Euler-Lagrange-2D-AA-1}\\
	N_{\text{T},j} &= \sum_{\mu} \frac{C_{j\mu} \left(\dot{\varphi}_{\text{T},j} - \dot{\varphi}_{\mu}\right)}{\left(2e\right)^{2}} + \frac{C_{\text{gT},j} \left(\dot{\varphi}_{\text{T},j} - 2e V_{\text{gT},j}\right)}{\left(2e\right)^{2}}\,,\label{eq:Euler-Lagrange-2D-AA-2}
\end{align}
and define the offset charges $N_{\text{g},\mu} \equiv C_{\text{g},\mu} V_{\text{g},\mu} / 2e$ and $N_{\text{gT},j} \equiv C_{\text{gT},j} V_{\text{gT},j} / 2e$ and the shifted charges $\widetilde{N}_{\mu} \equiv N_{\mu} + N_{\text{g},\mu}$ and $\widetilde{N}_{\text{T},j} \equiv N_{\text{T},j} + N_{\text{gT},j}$. Analogously to above, we construct a capacitance matrix $\mtrx{C}$ satisfying the equation $\vctr{Q} = \mtrx{C} \vctr{V}$ with the charge vector $\vctr{Q} = 2e \left(\widetilde{N}_{x}, \widetilde{N}_{y}, \widetilde{N}_{\text{T},1}, \widetilde{N}_{\text{T},2}\right)^\intercal$ and the voltage vector $\vctr{V} = \left(\dot{\varphi}_{x}, \dot{\varphi}_{y}, \dot{\varphi}_{\text{T},1}, \dot{\varphi}_{\text{T},2}\right)^\intercal / 2e$. Then we invert the matrix to find the kinetic part of the Hamiltonian via Legendre transformation $T = \vctr{Q}^{T} \mtrx{C}^{-1} \vctr{Q} / 2$. With Supplementary Equations~\eqref{eq:Euler-Lagrange-2D-AA-1} and \eqref{eq:Euler-Lagrange-2D-AA-2} we thus find the capacitance matrix
\begin{equation}
	\mtrx{C} =
	\begin{pmatrix}
		\mtrx{A} & \mtrx{B}\\
		\mtrx{B} & \mtrx{D}
	\end{pmatrix},
\end{equation}
with
\begin{equation}
	\mtrx{A} =
	\begin{pmatrix}
		C_{x,\text{tot}} & 0\\
		0 & C_{y,\text{tot}}
	\end{pmatrix},
\end{equation}
\begin{equation}
	\mtrx{B} =
	\begin{pmatrix}
		-C_{1x}           & -C_{2x}\\
		-C_{1y}           & -C_{2y}
	\end{pmatrix},
\end{equation}
and
\begin{equation}
	\mtrx{D} =
	\begin{pmatrix}
		C_\text{T1,tot} & 0\\
		0 & C_\text{T2,tot}
	\end{pmatrix},
\end{equation}
where we defined the total capacities $C_{\mu,\text{tot}} \equiv\sum_{j}C_{j\mu} + C_{\text{g},\mu}$ and $C_{\text{T},j,\text{tot}} \equiv C_{\text{gT},j} + \sum_{\mu}C_{j\mu}$. We denote the inverse of this matrix as
\begin{equation}
	\mtrx{C}^{-1} =
	\begin{pmatrix}
		\mtrx{U} & \mtrx{V}\\
		\mtrx{V}^\intercal & \mtrx{W}
	\end{pmatrix},
\end{equation}
which can be derived to be
\begin{equation}
	\mtrx{U} = \frac{1}{A - B\frac{1}{D}B^{T}} \,,
\end{equation}
\begin{equation}
	\mtrx{V} = -\frac{1}{A - B\frac{1}{D}B^{T}}B\frac{1}{D}\,,
\end{equation}
and
\begin{equation}
	\mtrx{W} = \frac{1}{D - B^{T}\frac{1}{A}B}\,.
\end{equation}
Defining the vectors \(\vctr{E}_\text{J} \equiv \left(E_{\mathrm{J},x}, E_{\mathrm{J},y}\right)^\intercal\), \(\vctr{\varphi} \equiv \left(\varphi_{x}, \varphi_{y}\right)^\intercal\), and \(\vctr{\widetilde{N}} \equiv \left(\widetilde{N}_{x}, \widetilde{N}_{y}\right)^\intercal\), we find for the full Hamiltonian
\begin{multline}
    H_{\text{AA}}^{(2)} = - \sum_{j=1}^{2} E_{\text{JT},j} \cos\left(\varphi_{\text{T},j}\right) + V^{(2)}_\text{CT}\\ - \vctr{E}_\text{J} \cdot \cos\left(\vctr{\varphi}\right)  + V_\text{C}^{(2)}\,,
\end{multline}
where we define the cosine of a vector component-wise $\Bigl[\cos\Bigl(\vctr{\varphi}\Bigr)\Bigr]_{\mu} := \cos\Bigl(\varphi_{\mu}\Bigr)$.
The islands forming the charge lattice contribute capacitively as
\begin{equation}
    V_\text{C}^{(2)} = 2\vctr{\widetilde{N}}^\intercal \mtrx{E}_\text{C} \vctr{\widetilde{N}},
\end{equation}
with the charging energy matrix $\mtrx{E}_\text{C}$. Furthermore, the capacitive energy contribution of the transmons is given by
\begin{multline}
    V^{(2)}_\text{CT} = \sum_{j=1}^{2} E_{\text{CT},j} \left(\widetilde{N}_{\text{T},j} + \vctr{\Lambda}_{j} \cdot \vctr{\widetilde{N}}\right)^{2}\\
    + E_{\text{CT},\text{cross}} \left(\widetilde{N}_{\text{T},1} + \vctr{\Lambda}_{1} \cdot \vctr{\widetilde{N}}\right) \left(\widetilde{N}_{\text{T},2} + \vctr{\Lambda}_{2} \cdot \vctr{\widetilde{N}}\right),
\end{multline}
with the transmon charging energies
\begin{equation}
    E_{\text{CT},j} = 2e^2 \frac{C_{x,\text{tot}} C_{y,\text{tot}} C_{\text{T}\overline{j},\text{tot}} - C_{\overline{j}x}^{2} C_{y,\text{tot}} - C_{\overline{j}y}^{2} C_{x,\text{tot}}}{\operatorname{det} \left[\mtrx{C}\right]}\,,
\end{equation}
($\overline{j} \neq j$), the cross coupling energy
\begin{equation}
    E_{\text{CT},\text{cross}} = 4e^2 \frac{C_{x,\text{tot}} C_{1y} C_{2y} + C_{y,\text{tot}} C_{1x} C_{2x}}{\operatorname{det} \left[\mtrx{C}\right]}\,,
\end{equation}
and the quasiperiodicity parameters
\begin{equation}
    \lambda_{j\mu} = \frac{C_{j\mu}}{C_{\mu,\text{tot}}}\,,
\end{equation}
forming the vectors \(\vctr{\Lambda}_j \equiv \left(\lambda_{jx}, \lambda_{jy}\right)^\intercal\). 
After quantizing the Hamiltonian, we assume tuning into an appropriate parameter regime where the transmon charges $\widehat{N}_{\text{T},j}$ have much faster dynamics than the island charges $\widehat{N}_{\mu}$ (the extensive discussion on this regime for the 1D case of the quasiperiodic capacitor in the main text can easily be extended to this multidimensional case). Consequently, we can derive an effective version of this Hamiltonian given by Eq.~(20) of the main text, characterized by a quasiperiodic capacitive behavior. The cross-coupling between the transmons $\sim E_{\text{CT},\text{cross}}$ is suppressed in the transmon limit, that is $E_{\text{CT},j} / E_{\text{JT},j} \ll 1$.

\section*{Supplementary Note 2}
\subsection*{Numerical evaluation of \texorpdfstring{$E_\text{S}$}{E\textunderscore S}}

The essence of the quasiperiodic nonlinear capacitor is to make use of the offset-charge dependence of the transmon ground state while requantizing this offset-charge living on a finite superconducting island. In this process we used the cosine approximation $-E_\text{S} \cos(2\pi N_\text{g})$ which is perfectly valid in the transmon regime ($E_\text{JT} \gg E_\text{CT}$) where the amplitude $E_\text{S}$ is exponentially suppressed; see Eq.~(6) of the main text. However, since we are interested in a measurable effect of this term, we suggest to go beyond the transmon regime, where $E_\text{JT} \sim E_\text{CT}$ or even $E_\text{JT} < E_\text{CT}$. This is outside of the regime where the above approximation is justified. This  motivates us to do a numerical evaluation of the exact amplitude $E_\text{S}$ as well as the harmonicity of the term; see Supplementary Figure~\ref{fig:ES_evaluation}.
\begin{figure}
    \centering
    \includegraphics[width=.95\linewidth]{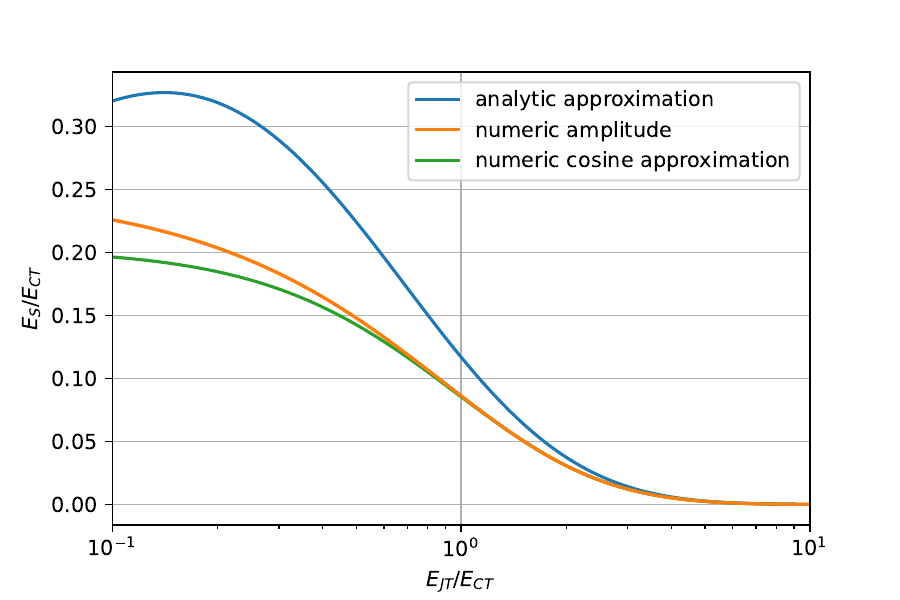}
    \caption{\textbf{Numeric evaluation of $E_\text{S}$.} We compare the analytic approximation from Eq.~(6) of the main text (blue line) with a numerically calculated amplitude (orange line) and the cosine component of the numerical result (green line). This is done as a function of $E_\text{JT} / E_\text{CT}$ in the transmon regime ($E_\text{JT} \gg E_\text{CT}$) and beyond.
    }
    \label{fig:ES_evaluation}
\end{figure}
To test the latter, we compute the Fourier coefficient of the numeric eigenvalue with the same frequency as the analytic term and compare it to the full numeric amplitude.
We find that, while we overestimate $E_\text{S}$ a bit outside the transmon regime, the analytic result still captures roughly the correct magnitude. Furthermore, the anharmonicity stays rather small, which yields only a small contribution of phase cotunneling events.

\section*{Supplementary Note 3}
\subsection*{Numerical evaluation of the Dirac spectrum}

To derive the Dirac Hamiltonian (see Eq.~(10) of the main text) the Hilbert space was reduced to only a single pseudo-spin degree of freedom and two externally controlled parameters. First we got rid of the degrees of freedom of the auxiliary transmon of the QPNC and afterwards we used Bloch's theorem to find the final 2D low-energy basis with Dirac-like dynamics. To affirm the validity of this process and the resulting energy profile, we numerically evaluate the Hamiltonian before and after analytically solving the dynamics of the auxiliary transmon (see Supplementary Figure~\ref{fig:Dirac_point}) and compare it to the expected Dirac spectrum. 
\begin{figure}
    \centering
    \includegraphics[width=\linewidth]{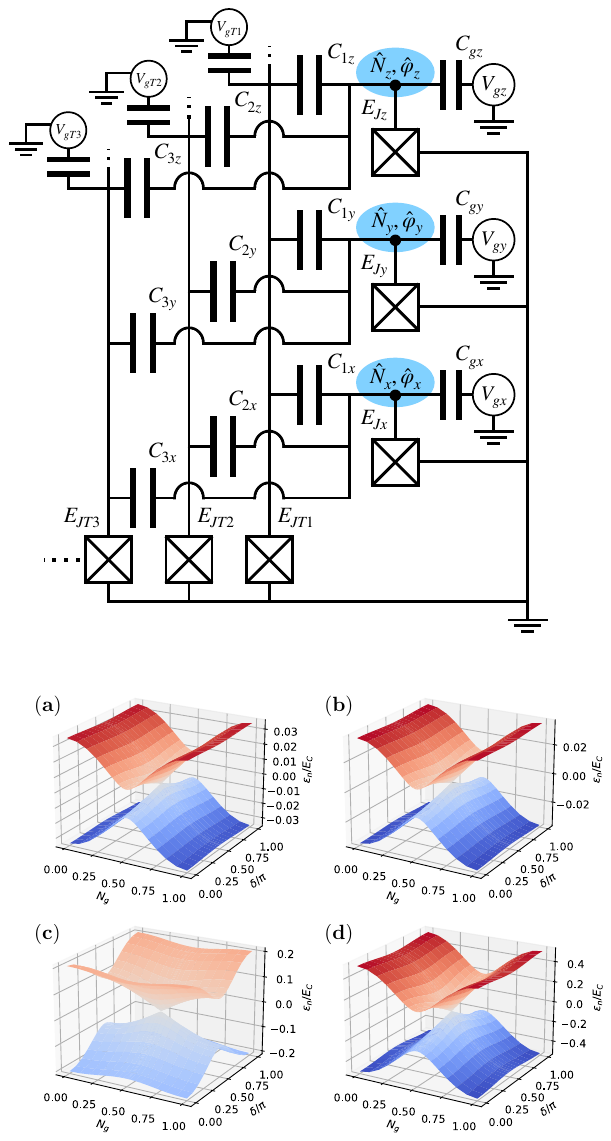}
    \caption{\textbf{Tests of the approximation procedure arriving at the Dirac cones.} We compare the eigenspectrum $\epsilon_n$ of the Dirac circuit (see Fig.~1(c) of the main text) including explicitly the auxiliary transmon dynamics [shown in panels (a, c)] and with the effective treatment of the quasiperiodic nonlinear capacitor, i.e., where we assume that the auxiliary transmon remains in the ground state (b, d). In (a, b) we used the parameters $E_\text{J} = 0.01 E_\text{C}$, $E_\text{2J} = 20 E_\text{C}$, and in (a) $E_\text{JT} = 4 E_\text{CT} = 16 E_\text{C}$, and in (b) the equivalent $E_\text{S} \approx 0.024 E_\text{C}$. 
    We find a very good agreement of the spectra of the two Hamiltonians. In (c) we choose a lower $E_\text{JT} = E_\text{CT}$, and respectively $E_\text{S} \approx 0.47 E_\text{C}$ in (d), while tuning up $E_\text{J} = 0.2 E_\text{C}$ accordingly in both (c, d). Here, neglecting the auxiliary transmon dynamics was not justified (since (c) and (d) give quantitatively different spectra). However, the discrepancy can be easily understood by means of excitations in the auxiliary transmon, see also main text.
    }
    \label{fig:Dirac_point}
\end{figure}
We perform this comparison for two sets of parameters. First, we choose a parameter regime where our approximations in the main text are expected to hold; see Supplementary Figures~\ref{fig:Dirac_point}(a, b). Here we see indeed a very good qualitative and quantitative agreement with the expected Dirac cone spectrum, i.e., the eigenvalues of the Hamiltonian in Eq.~(10) of the main text. Then, in Supplementary Figures~\ref{fig:Dirac_point}(c, d), we plug in values for which the approximations cannot be expected to hold. Surprisingly, we see that nonetheless, the Dirac point feature is still preserved in this regime. But the agreement is now only qualitative, and no longer quantitative between Supplementary Figures~\ref{fig:Dirac_point}(c) and (d). In particular, the amplitude of the energy spectrum with respect to $N_\text{g}$ is reduced in the full calculation of Supplementary Figure~\ref{fig:Dirac_point}(c). This is due to the fact that here, the energy of the Cooper-pair cotunneling junction, $E_\text{2J}=20 E_\text{C}$, is chosen very large compared to the excitation gap of the auxiliary transmon, which, for the chosen parameters, scales as $\sim 2\sqrt{E_\text{CT} E_\text{JT}} = 8 E_\text{C}$. As pointed out in the main text, this leads to excitations in the auxiliary transmon state, effectively reducing $E_\text{S}$. This fully explains the difference between the two figures.
\linebreak

\section*{Supplementary Note 4}
\subsection*{Effect of a \texorpdfstring{$\delta \lambda$}{lambda}-detuning on the Dirac spectrum}

In the Dirac circuit, the large $E_\text{2J}$ element stabilizes phase slips of exactly $\pi$, which enter in the effective Dirac Hamiltonian via the $-E_\text{S} \cos(\pi \widehat{N})$ term. This allows us to suppress the linear capacitance with a $\cos(2 \widehat{\varphi})$-term without suppressing the nonlinear capacitance. As we already discussed in the main text, a detuning $\delta \lambda$ from the theoretically optimal value, $\lambda = 1/2 + \delta \lambda$, does not influence the spectrum within the 2D low-energy basis. In addition, a numerical analysis of the full Hamiltonian shows that, to the contrary, a small detuning from $\lambda = 1/2$ can surprisingly be beneficial. Namely, a specific finite detuning $\delta \lambda$ will actually exactly close the gap and thus counteract the perturbing effects of other imperfections like the influence of higher-energy states.

\section*{Supplementary Note 5}
\subsection*{Classical current-noise spectrum in more details}
We consider an adiabatic current response (see Eq.~(12) of the main text) due to classical offset charge noise, as described in the main text. As the two contributions to the response we find the Josephson current $I_0 (t) = 2e\, \partial_\delta \epsilon_n$ and a linear correction term $I_1 (t) = -2e \dot{N}_\text{g} \mathcal{B}_0 [\delta, N_\text{g} (t)]$ proportional to the Berry curvature $\mathcal{B}_0$.
Exploiting the periodicity of the current in $N_\text{g}$, we now cast the different contributions into a discrete Fourier series, $I_0 = \sum_{n \in \mathbb{Z}} i_n \exp(\mathrm{i} \pi n N_\text{g})$ and $2e \mathcal{B}_0 = \sum_{n \in \mathbb{Z}} b_n \exp(\mathrm{i} \pi n N_\text{g})$ such that $I_1 = \partial_t \sum_{n \in \mathbb{Z}} \mathrm{i} b_n \exp(\mathrm{i} \pi n N_\text{g}) / (\pi n)$. Thus we can straightforwardly calculate the noise spectrum and find two contributions, one from the Josephson current (with subscript J) and a Berry curvature contribution (with subscript $\mathcal{B}$)
\begin{equation}\label{eq:current-noise-spectrum}
	S\left(\omega\right) = S_\text{J}\left(\omega\right) + S_{\mathcal{B}}\left(\omega\right),
\end{equation}
with
\begin{equation}
	S_\text{J}\left(\omega\right) = 4T \sum_{n > 0} \frac{\left(\pi n\right)^2 \left|i_n\right|^2}{\left(\pi n\right)^4 + T^2 \omega^2}\,,
\end{equation}
and
\begin{equation}
	S_{\mathcal{B}}\left(\omega\right) = 4T \sum_{n > 0} \frac{\omega^2 \left|b_n\right|^2}{\left(\pi n\right)^4 + T^2 \omega^2}\,.
\end{equation}
The total current noise $S$ and its two contributions $S_\text{J}$ and $S_{\mathcal{B}}$ are shown for the Dirac system and its gapped version in Fig.~3 of the main text.

When considering the phase $\delta$ at a value away from the degeneracies at $\pi/2$ and $3\pi/2$, we can identify the dominant Fourier coefficients $i_{m_i}$ and $b_{m_b}$. We thus find an approximate Lorenz curve of height $S_0 = 4T |i_{m_i}|^2 / (\pi m_i)^2$ for $S_\text{J}$, and with an asymptotic behavior of $S_\text{J}(\omega) \rightarrow \omega^{-2} \left(2 \pi m_i\right)^2 \left|i_{m_i}\right|^2 / T$. For $S_{\mathcal{B}}$ we find an inverted Lorenz curve, that is, a constant $S_\infty$ minus a Lorenz curve of the same height $S_\infty = 4 |b_{m_b}|^2 / T$, which gives us the asymptotic value of the spectrum for high frequencies $\left|\omega\right| \rightarrow \infty$.
Importantly, the asymptotic value vanishes $S_\infty = 0$ in the Dirac system since the Berry curvature (and thus every $b_{n}$) is zero, in contrast to a gapped system; see Fig.~3 of the main text. Therefore, we can probe whether or not the system is gapped by measuring this offset value $S_\infty$. With this in mind let us briefly discuss the behavior of the current noise contributions for a system that is possibly gapped.

To compare the two contributions, $S_\text{J}$ and $S_{\mathcal{B}}$, it is useful to consider the parameter $\beta \equiv S_{\infty} / S_0 = (m_i \pi)^2 |b_1|^2 / |i_2|^2 T^2$. However, note that the two Lorenz curves in general have different widths due to different dominant Fourier coefficients of the current contributions contributing at different frequency values $m_i$ and $m_b$, as discussed above. Hence we find a non-monotonous noise spectrum $S(\omega)$ for values of $\beta$ close to 1; see Fig.~3(a) of the main text.
Considering the symmetric case of $E_\text{S} = E_\text{J} \equiv E$ and a small gap $\Delta^2 \ll E^2$, we can estimate the scaling of the noise spectrum contributions as $S_0 \sim T E^2$ and $S_{\infty} \sim \Delta^2 / T E^2$ such that $\beta \sim \Delta^2 / T^2 E^4$.

\end{document}